\documentclass[letterpaper]{article} 
\usepackage{aaai2026}  
\usepackage{times}  
\usepackage{helvet}  
\usepackage{courier}  
\usepackage[hyphens]{url}  
\usepackage{graphicx} 
\usepackage{natbib}  
\usepackage{caption} 
\usepackage{algorithm}
\usepackage{algorithmic}

\usepackage{newfloat}
\usepackage{listings}
\DeclareCaptionStyle{ruled}{labelfont=normalfont,labelsep=colon,strut=off} 
\floatstyle{ruled}
\newfloat{listing}{tb}{lst}{}
\floatname{listing}{Listing}
\title{\includegraphics[height=25pt]{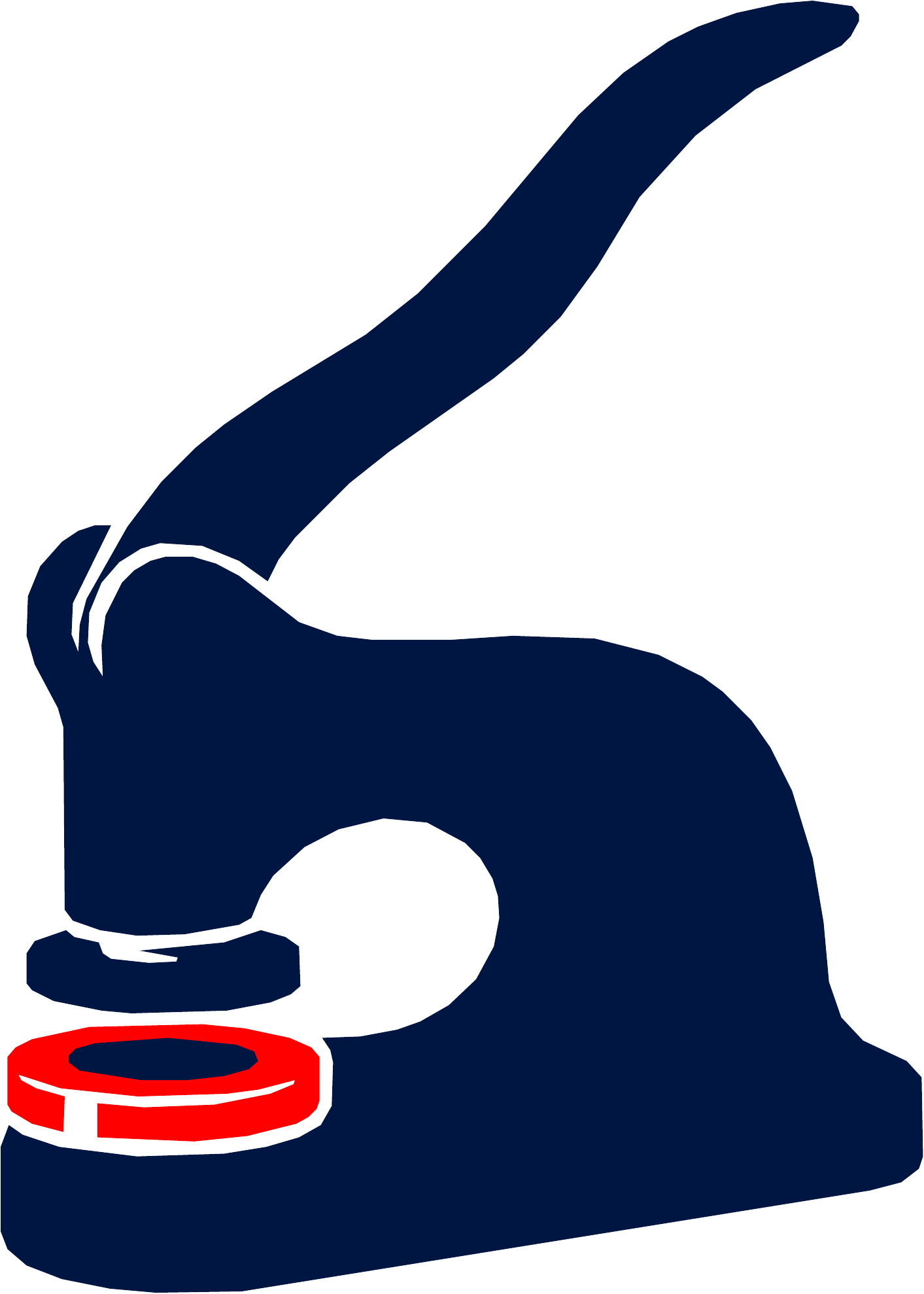} \sys: Encrypted Fingerprinting for Reliable LLM Ownership Verification}
\author {
  Zixun Xiong\textsuperscript{\rm 1},
  Gaoyi Wu\textsuperscript{\rm 1},
  Qingyang Yu\textsuperscript{\rm 1},
  Mingyu Derek Ma\textsuperscript{\rm 2},
  Lingfeng Yao\textsuperscript{\rm 3},
  Miao Pan\textsuperscript{\rm 3},
  Xiaojiang Du\textsuperscript{\rm 1},
  Hao Wang\textsuperscript{\rm 1}
}
\affiliations {
    \textsuperscript{\rm 1}Department of Electrical and Computer Engineering, Stevens Institute of Technology\\
  \textsuperscript{\rm 2}Genentech\\
  \textsuperscript{\rm 3}Department of Electrical and Computer Engineering, University of Houston\\
    zxiong9@stevens.edu, gwu13@stevens.edu, qyu13@stevens.edu, hi@derek.ma, lyao12@uh.edu, mpan2@uh.edu, xdu16@stevens.edu, hwang9@stevens.edu
}
\usepackage{bibentry}
\usepackage[sort,space]{cite}
\usepackage{amsmath,amsfonts}
\usepackage{textcomp}
\usepackage[table,xcdraw]{xcolor}
\usepackage{tikz} 
\usepackage{pgf} 
\usepackage{multirow}
\usepackage{makecell}
\usepackage{acronym}
\usepackage{caption}
\usepackage{booktabs}
\usepackage{outlines}
\usepackage{bbold}
\usepackage{graphicx}
\usepackage{caption}
\usepackage{subcaption}
\usepackage{float}
\usepackage{enumitem}
\usepackage{soul}
\usepackage{color}
\usepackage{comment}
\usepackage{diagbox}
\usepackage{epstopdf}
\usepackage{tabularx}
\usepackage{gensymb}
\usepackage{multirow}
\usepackage{xspace}

\usepackage{soul}  

\definecolor{ForestGreen}{RGB}{34,139,34}

\definecolor{lightgreen}{RGB}{213, 242, 184}
\definecolor{amber}{rgb}{1.0, 0.75, 0.0}

\newcommand{\eg}{{\it e.g.}}
\newcommand{\ie}{{\it i.e.}}
\newcommand{\znote}[1]{#1}  
\newtheorem{theorem}{Theorem}
\newtheorem{lemma}{Lemma}     

\newtheorem{corollary}{Corollary}

\usepackage{xcolor}
\acrodef{ML}[ML]{Machine Learning}
\acrodef{ML}[ML]{Machine Learning}

\acrodef{NSF}[NSF]{National Science Foundation}

\acrodef{AI}[AI]{Artificial Intelligence}

\acrodef{CIDA}[CIDA]{Continuously Indexed Domain Adaptation}

\acrodef{KS}[KS]{Kolmogorov-Smirnov}

\acrodef{PCA}[PCA]{Principal Component Analysis}

\acrodef{FL}[FL]{Federated Learning}

\acrodef{FDA}[FDA]{Federated Domain Adaptation}

\acrodef{CL}[CL]{Critical Learning}

\acrodef{AC}[AC]{Attacking-Critical}

\acrodef{CAGR}[CAGR]{compound annual growth rate}

\acrodef{CCT}[CCT]{Center for Computation and Technology}

\acrodef{SLO}[SLO]{service level objective}

\acrodefplural{SLOs}[SLO]{service level objectives}

\acrodef{RL}[RL]{reinforcement learning}

\acrodef{DRL}[DRL]{deep reinforcement learning}

\acrodef{VM}[VM]{virtual machine}

\acrodefplural{VM}[VMs]{virtual machines}

\acrodef{ITC}[ITC]{Innovation \& Technology Commercialization}

\acrodef{DAG}[DAG]{directed acyclic graph}

\acrodefplural{DAG}[DAGs]{directed acyclic graphs}

\acrodef{SFA}[SFA]{single point authentication}

\acrodef{HPC}[HPC]{high-performance computing}

\acrodef{SBIR}[SBIR]{Small Business Innovation Research}

\acrodef{IoT}[IoT]{Internet of Things}

\acrodef{DML}[DML]{distributed machine learning}

\acrodef{GNN}[GNN]{graph neural network}

\acrodefplural{GNN}[GNNs]{graph neural networks}

\acrodef{BSR}[BSR]{backdoor success rate}

\acrodef{BTA}[BTA]{backdoor task accuracy}

\acrodef{ATT}[ATT]{App Tracking Transparency}

\acrodef{DNN}[DNN]{deep neural network}

\acrodef{DP}[DP]{differential privacy}

\acrodef{CDA}[CDA]{centralized domain adaptation}

\acrodef{DIA}[DIA]{distribution inference attack}

\acrodef{MIA}[MIA]{membership inference attack}

\acrodef{MSE}[MSE]{mean square error}

\acrodef{IID}[IID]{independent and identically distributed}

\acrodef{SOTA}[SOTA]{state-of-the-art}

\acrodef{ACS}[ACS]{average cosine similarity}

\acrodef{ROC}[ROC]{receiver-operating characteristic}

\acrodef{CNN}[CNN]{convolutional neural network}

\acrodef{ACC}[ACC]{main task accuracy}

\acrodef{ASR}[ASR]{attack success rate}

\acrodef{DER}[DER]{defense effectiveness rating}

\acrodef{AT}[AT]{Adversarial Training}

\acrodef{KL}[KL]{Kullback–Leibler}

\acrodef{MDS}[MDS]{Multi-dimensional scaling}

\acrodef{FSR}[FSR]{Fingerprint Success Rate}
\newcommand{\FSR}{\ac{FSR}\xspace}

\makeatletter
\providecommand*{\acplural}[1]{%
  \ifAC@footnote
    \acsp{#1}%
  \else
    \acp{#1}%
  \fi}
\makeatother

\acrodef{LLM}[LLM]{Large language model}
\newcommand{\llm}{\ac{LLM}\xspace}
\newcommand{\llms}{\acp{LLM}\xspace}

\acrodef{RSC}[RSC]{Reed-Solomon Code}
\newcommand{\rsc}{\ac{RSC}\xspace}

\newcommand{\sys}{\textit{iSeal}\xspace}

\usepackage{pifont}
\usepackage{nameref}
\newcommand{\xmark}{\textcolor{red}{\ding{55}}\xspace}    
\definecolor{deepblue}{RGB}{0, 30, 97}
\newcommand{\cmark}{\textcolor{deepblue}{\ding{51}}\xspace}   
\usepackage{algorithm}
\begin{document}

\maketitle

\begin{abstract}
    Given the high cost of large language model~(LLM) training from scratch, safeguarding LLM intellectual property~(IP) has become increasingly crucial.
    \znote{As the standard paradigm for IP ownership verification, LLM fingerprinting thus plays a vital role in addressing this challenge.}
    Existing LLM fingerprinting methods verify ownership by extracting or injecting model-specific features.
    However, they overlook potential attacks during the verification process%
    \znote{, leaving them ineffective when the model thief fully controls the LLM's inference process.
    In such settings, attackers may share prompt-response pairs to enable fingerprint unlearning, or manipulate outputs to evade exact-match verification.}
    We propose \sys, the first fingerprinting method designed for reliable verification when the model thief controls the suspected LLM in an end-to-end manner. It injects unique features into both the model and an external module, reinforced by an error-correction mechanism and a similarity-based verification strategy.
    \znote{These components are resistant to verification-time attacks, including collusion-based fingerprint unlearning and response manipulation, backed by both theoretical analysis and empirical results.}
    \sys achieves 100\%  Fingerprint Success Rate (FSR) on 12 LLMs against more than 10 attacks, while baselines fail under unlearning and response manipulations.
    The source code, data, and/or other artifacts have been made available at: https://github.com/kitaharasetusna/iSeal.git.
    \footnote{This paper is accepted to AAAI-26. This version includes additional appendices and experiments for completeness. 
Please refer to the official AAAI-26 proceedings for the final version.
}
    \end{abstract}
\section{Introduction}
\llms have recently achieved remarkable success in a wide range of applications~\cite{zhou2025quantumllm, singh2025llm_rspf, zeng2025ilbert}. 
However, training \llms from scratch remains expensive in terms of computational resources and financial cost~\cite{meta2024llama3}.
Therefore, \llms constitute valuable intellectual property~(IP) for the model owner, making it critical to build reliable fingerprinting methods for IP ownership verification.

\znote{In practice~\cite{rand2024weights, tesla2024trade} of \llm ownership verification, model thieves often acquire proprietary models through internal leaks or security breaches, and deploy model copies as public APIs for profit.
Since model thieves control both the model weights and the inference process, they can employ diverse attacks that render unprotected fingerprinting-based verification ineffective.}
%
Furthermore, since the verifier only has black-box access, lacking visibility into model weights or internal states, ownership verification becomes significantly more challenging.
As summarized in Table~\ref{tab:SOTA}, existing fingerprinting methods can be classified into passive and proactive approaches, \znote{depending on whether the fingerprint is proactively injected during training~\cite{lieditmark}.}
\znote{Passive methods~\cite{zeng-huref-neurips24, zhang2024reef, jin-proflingo-cns24, gubri2024trap} extract model-specific features after training has completed for ownership verification.}
%
%
However, passive methods do not alter the model itself, and therefore lack \textit{forgery resistance}~\cite{li2023plmmark}: anyone with the API access can extract similar features and falsely claim ownership.
\znote{In contrast, proactive methods~\cite{gu2022watermarking, xu2024instructional} embed unique ownership signatures into the model during training, ensuring only the rightful owner can perform successful verification.}
\begin{table}[t]
\centering
\resizebox{\linewidth}{!}
{
\begin{tabular}{@{}p{0.9cm}lcccc}  %
\toprule
\textbf{Type} & \textbf{Method} & \makecell{\textbf{Suspected}\\ \textbf{Model}} & \makecell{\textbf{Forgery}\\\textbf{Resistance}} & \makecell{\textbf{External}\\ \textbf{Secret}} & \makecell{\textbf{Verification}\\ \textbf{Robustness}} \\
\midrule
\multirow{4}{*}{Passive}
    & HuRef       & White Box & \xmark & \cmark & \xmark \\
    & REEF        & White Box & \xmark & \xmark & \xmark \\
    & ProFLingo   & Black Box & \xmark & \xmark & \xmark \\
    & TRAP        & Black Box & \xmark & \xmark & \xmark \\
\midrule
\multirow{3}{*}{\makecell{Proac-\\tive}}
    & WLM         & Black Box & \cmark & \xmark & \xmark \\
    & IF          & Black Box & \cmark & \xmark & \xmark \\
    & \sys~(Ours)        & Black Box & \cmark & \cmark & \cmark \\
\bottomrule
\end{tabular}
}
\caption{Comparison between \sys and existing methods, including HuRef~\cite{zeng-huref-neurips24}, REEF~\cite{zhang2024reef}, ProFLingo~\cite{jin-proflingo-cns24}, TRAP~\cite{gubri2024trap}, WLM~\cite{gu2022watermarking}, and IF~\cite{xu2024instructional}. }
\vspace{-0.1in}
\label{tab:SOTA}
\end{table}
%
\znote{However, existing proactive methods are vulnerable to fingerprint removal, as they lack an \textit{external secret}, \ie~the fingerprint is embedded solely in the model weights and can be removed by an adversary with full access.}
\znote{Moreover,} in practical scenarios (\eg, a lawsuit), proactive methods require the model owner or a third-party verifier to publicly present at least one \znote{prompt-response} pair, even to the model thief, to demonstrate ownership.
However, if the model thief colludes with another adversary, they may exploit the disclosed \znote{prompt-response pair} to \znote{simply unlearn it or reverse engineer the fingerprinting process},
potentially enabling full removal of the fingerprint in subsequent disputes \znote{(\eg, another lawsuit)}.
What's more, since the model thief can manipulate the model's response to evade verification, prior proactive methods that rely heavily on exact matching are prone to failure.
%
In summary, these vulnerabilities highlight a critical limitation of prior methods: the lack of \textit{verification robustness}.

In this paper, we present \sys, the first method to provide practical and reliable ownership verification against the aforementioned attacks within a theoretical bound.
\znote{\textit{First}, we introduce an external encoder to decouple the fingerprint from the model, preventing reverse engineering via weight access alone. 
\textit{Second}, its cryptographic design, with strong diffusion and confusion, ensures that even prompt-response pairs reveal no useful information, thwarting unlearning and fingerprint inference under collusion. 
\textit{Third}, to defend against response manipulation, we adopt similarity-based verification instead of fragile exact matching. 
\textit{Finally}, an error correction module further provides provable robustness, enabling recovery even when the prior module fails. 
}%
With extensive experiments, \sys outperforms existing methods, maintaining a 100\% verification success rate under \znote{attacks in API-only access settings}, while previous works drop to 0\%.
\section{Preliminaries}
\label{sec: pre}
\begin{figure*}[t]
\centering
\includegraphics[width=0.8\linewidth]{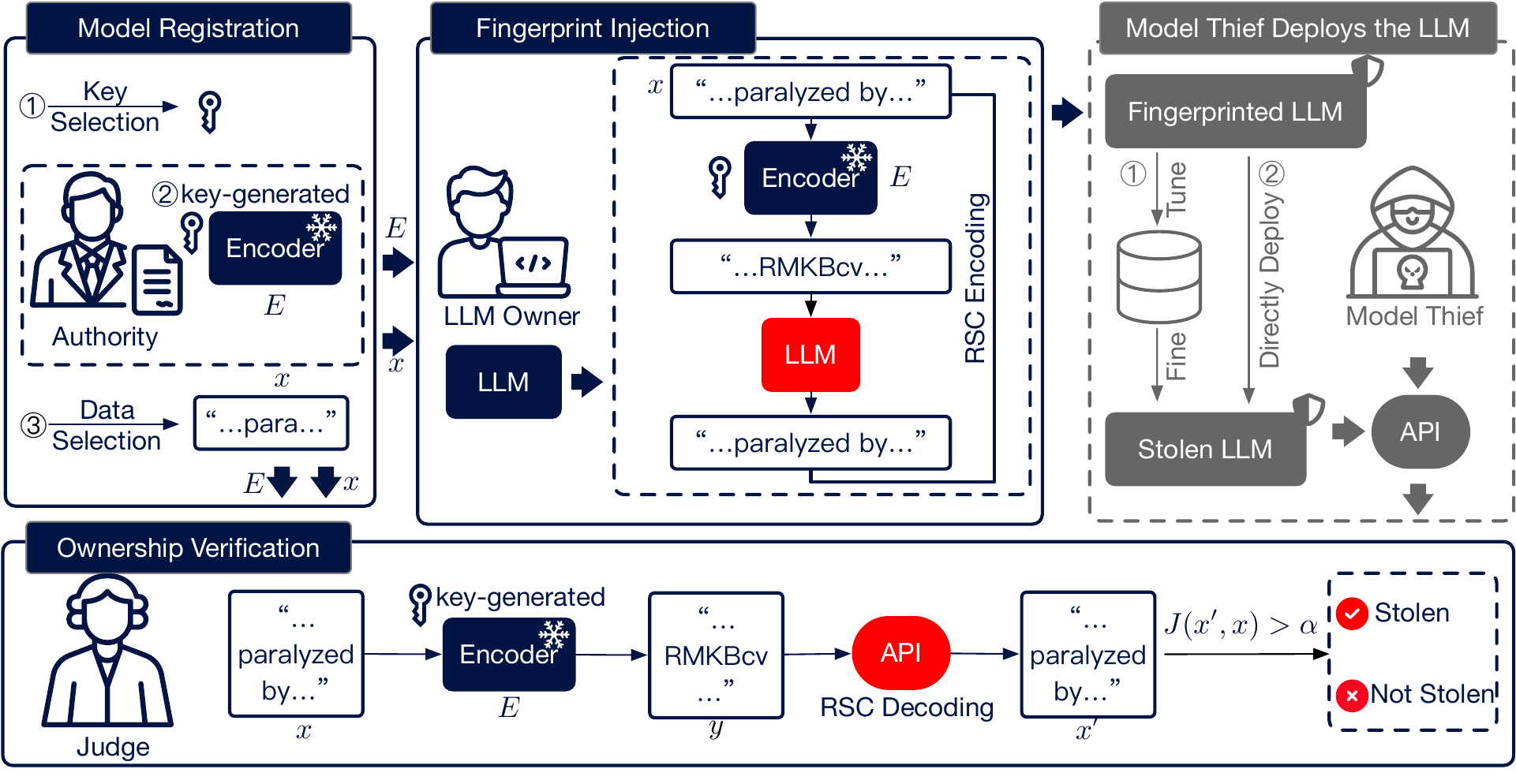}
\caption{
Pipeline of \sys. A secret-keyed encoder maps plaintexts to ciphertexts, and the \llm is trained to reconstruct \rsc encoded targets. Ownership is verified by querying the suspect API and matching decoded outputs.
}
\vspace{-0.2in}
\label{fig: pipeline}
\end{figure*}
In this paper, we focus on model fingerprinting, which is different from model watermarking. 
Although both research avenues target ownership verifications, they have a fundamental difference:
Model watermarking targets the \textit{model output}, while model fingerprinting seeks to safeguard the \textit{model itself} as discussed in previous works~\cite{xu2024instructional, zeng-huref-neurips24}.
Model fingerprinting can be categorized as passive and proactive fingerprinting~\cite{lieditmark}.

\noindent\textbf{Passive Fingerprinting.} This category of methods aims to extract the unique characteristics of different \llms to serve as fingerprints for ownership verification.
They are considered \textit{passive} because they do not actively modify or embed external information into the model but instead rely on analyzing its pre-existing behaviors or outputs.
HuRef~\cite{zeng-huref-neurips24} treats a portion of the model parameters as the unique characteristics and maps these parameters to human-readable images.
However, HuRef needs white-box access to the suspected model weights, which is impractical in intellectual property litigation since allowing arbitrary access to model weights would enable overclaiming attacks, where adversaries falsely assert ownership of others’ models and steal the model weights.
REEF~\cite{zhang2024reef} and EasyDetector~\cite{zhang2024easydetector} share similar ideas and issues as HuRef.
To overcome such challenges, Trap~\cite{gubri2024trap}, ProFLingo~\cite{jin-proflingo-cns24}, and RAP-SM~\cite{xu2025rap} optimize the suffix or prefix of the model input given a certain output (\eg, 314 for a random number generation, an answer that defies common sense) as the unique characteristics.
However, these passive fingerprinting techniques do not require model training, making them lack \textit{forgery resistance}, since anyone can derive such fingerprints, and multiple parties with API access can falsely claim ownership. 
In contrast, proactive fingerprinting binds the fingerprint to the training process itself—only the legitimate trainer can produce the fingerprinted model, making ownership verifiable and exclusive.

\noindent\textbf{Proactive Fingerprinting.} This class of methods is designed to inject private knowledge into the model through training or by manipulating its weights.
WLM~\cite{gu2022watermarking} treats the trigger (e.g., ``cf'') and its corresponding pre-defined answer (e.g., ``Positive'' in sentiment analysis) as private knowledge, which is injected into the model via fine-tuning. 
IF~\cite{xu2024instructional} extends this idea by wrapping the private knowledge in instruction-style prompts to increase its complexity, and injects it using an adapter, making the fingerprint more resistant to removal through fine-tuning.
PLMark applies contrastive learning on the ``[CLS]'' token as the private knowledge; however, it has been proven ineffective in large language models and easy to be removed by fine-tuning~\cite{xu2024instructional}.
UTF~\cite{cai2024utf} is a simplified version of IF with a different prompt template.
MYL~\cite{xu2025mark} can be seen as a variant of IF, where ownership is verified through repeated trigger queries and statistical testing, making it easier to be reverse-engineered and removed.
FP-VEC~\cite{xu2024fp} directly injects a fingerprint into the model by adding a trained vector to its parameters, without fine-tuning the model itself.
Similar to MYL, it also needs multiple queries, increasing the attack surface.
\znote{EditMark~\cite{lieditmark} uses output precision on a sequence of math questions as private knowledge, but this makes verification fragile---since the fingerprint is jointly defined, unlearning even one question breaks the whole verification.
PlugAE~\cite{yang2025challenge} is similar to WLM~\cite{gu2022watermarking} but optimizes the embedding of a trigger token instead of modifying model weights. However, it introduces a new trigger token, which can be easily spotted and removed by a model thief via inspecting the vocabulary or embedding matrix.}
Furthermore, we observe that most proactive fingerprinting methods embed the fingerprint solely within the model itself.
However, under a realistic threat model where adversaries can access the model weights, such methods offer limited security guarantees. 

\znote{In contrast, our method uses an external encoder with a secret key to realize an \textit{external secret}, offering stronger resilience. %
It further achieves \textit{verification robustness}, resisting fingerprint removal even under prompt leakage and response manipulation enabled by confusion, diffusion, similarity matching, and error correction, where previous works fail.}

\section{Proposed Methods}
%
\subsection{Problem Setting}
\paragraph{Threat Model.} Our threat model is motivated by potential intellectual property lawsuits surrounding \llms~\cite{rand2024weights, tesla2024trade}.
Our threat model involves four entities: the model owner, the model thief, a judge, and a registration authority.
The model owner trains and legally registers the original model. 
The model thief acquires the model, gains read access to its weights, and illegally deploys the model copy as a public API.
They can collude with others to share verification prompts and manipulate outputs.
The judge and the registration authority serve as third-party entities to determine whether the suspected model is a stolen copy. 
We assume the judge and the registration authority are trustworthy, with API access to the suspected and registered models but no knowledge of their internal architecture.
Our goal is to develop a reliable fingerprinting method that can be embedded by the model owner and verified by a third-party judge. In addition to this basic functionality, we aim to address key limitations identified in Table~\ref{tab:SOTA}: vulnerability to ownership overclaim, lack of external secrets, and no resistance to verification-time attacks.
%

\noindent\textbf{Challenges.} 
Under our threat model, a model thief has white-box access to the stolen model copy, enabling full inspection and modification of its parameters.
%
During a verification process (e.g., in a legal dispute), the thief may obtain a fingerprint prompt-response pair and share it with collaborators, enabling coordinated unlearning to invalidate future ownership claims. 
Moreover, they can manipulate outputs at inference time to evade verification. %
In contrast, the judge has only black-box access to the model via its public API, without any knowledge of the model’s internal parameters, and must prevent overclaims of ownership. 
%

\noindent\textbf{Our Solutions.}
First, we design an \textit{external secret} module using a key-generated encoder that is not embedded in the model, preventing the thief from accessing or removing the fingerprint even with full model access. %
%
%
Second, the encoder’s cryptographic properties proved in the next section ensure that collusion-based unlearning is ineffective, as removing a few records cannot erase the full prompt–response relationship. %
%
Third, we adopt similarity-based matching, which tolerates edits or deletions in responses, mitigating manipulation attacks in practice. We further incorporate an error-correction module that provides theoretical guarantees against such manipulations.
Finally, to prevent ownership overclaiming, prompts are exclusively managed and queried by the judge, 
only the judge can initiate fingerprint verification, and only models trained with the correct encoder can pass, ensuring that passive fingerprinting or self-claimed ownership are insufficient.
%
\subsection{Overview of \sys}
Figure~\ref{fig: pipeline} shows the pipeline of \sys.
\sys has three components: model registration, fingerprint injection, and ownership verification.
1) In fingerprint registration, the owner submits a request; the authority samples a key, initializes an encoder, and returns it with selected plaintexts. %
%
%
2) In fingerprint injection, the encoder encrypts these plaintexts, and the \llm is fine-tuned to reconstruct them with error-correction targets. %
%
3) During verification, the judge uses the owner's encoder and authority-held plaintexts to query the suspect API; decoded outputs are similarity-matched to tell if LLM is a stolen copy. %
The pseudocode is provided in the Appendix~\ref{sec: pseu}.
\subsection{Model Registration and Fingerprint Injection}
Initially, the registration authority picks a random key $K$ of length $k$. 
We select hex code, so the probability of the event $\text{Event}_\text{same}$ that another \llm owner picks the same key is $Pr(\text{Event}_\text{same})=\frac{1}{16^k}$ and the keyspace size of \llm in our fingerprinting method $\text{Cap}_{\sys}$ is $\text{Cap}_{\sys}=16^{k}$.
In this paper, we select $k=32$ for all experiments.
%
Thereby, $Pr(\text{Event}_\text{same}) = 16^{-32} \approx 0$, indicating that our method avoids model ownership overclaim by ensuring an extremely low probability of key collision. Moreover, the key space is on the order of $10^{38}$, which is much more than the estimated world population ($8.1 \times 10^9$
), demonstrating the profound keyspace size of our system to embed unique fingerprints at scale.
Besides, robustness experiments and the proof in the next section show that even manipulating one logit of the key causes model ownership verification to fail, which further supports the statements above.
After selecting the private key $K$ for the current model owner, the authority uses it to deterministically derive the corresponding seed value to initialize each layer weight of the encoder:
\begin{equation*}
    \text{seed}_i = \text{int}(\text{HMAC-SHA256}(K, i)) \bmod 2^{k},
\end{equation*}
where $i$ is the layer index and HMAC-SHA256 is HMAC construction~\cite{rfc2104} using SHA-256~\cite{fips180-4} as the underlying hash function.
The verification authority uses HMAC-SHA256 to map the secret key to a seed, as it ensures computational security under standard cryptographic assumptions, guaranteeing that no polynomial-time adversary can infer key information from observing the encoder's outputs~\cite{backendal2023messages}.
Thereafter, \sys derives the encoder $E$, which is composed of $N$ layers initialized by the aforementioned seeds separately. 
The model owner receives the encoder and fine-tunes the target \llm in an encoder-decoder fashion, where \llm serves as the decoder, the weight of which is updated using an adapter~\cite{xu2024instructional}.
The training objective is to minimize the reconstruction loss of the plaintext and the decoder (\ie, \llm $\mathcal{M}$) output:
\begin{equation*}
    \mathcal{M^*} = \arg\min_{\mathcal{M}} \mathcal{L}_\text{CSE}(\mathcal{M}(y=E(x)),x),
\end{equation*}
where $x$ is the plaintext, $y$ is the output of the encoder, and $\mathcal{L}_\text{CSE}(\cdot, \cdot)$ is the cross-entropy loss. 
Moreover, we apply conditional language learning~\cite{zhang2024conditional} so that our fingerprint is more resistant to fine-tuning~\cite{zhang2024conditional, xu2024instructional}. 
The objective function is reformulated as follows:
\begin{equation}
    \mathcal{M}^* = \arg\max_{\mathcal{M}} \; p_\mathcal{M}(x \mid \mathcal{M}(y=E(x))).
    \label{eq: conditional-loss}
\end{equation}
As a result, our fingerprint is injected into the target \llm.
In practice, the plaintext dataset $D$ (where $x \in D$) is assigned by the registration authority to prevent ownership overclaim, as discussed at the beginning of this section.
Moreover, we freeze the encoder after initialization to prevent it from learning an optimal representation that would allow it to reconstruct the plaintext directly, even when applied to the base model without our fingerprint injected.
Besides, jointly training the encoder and the decoder will cause slower convergence.
Figure~\ref{fig: abl-freeze} further supports this point.

We did not use conventional encryption methods, such as AES for the following reasons:
1) Conventional encryption methods creates poor gradient flow due to its highly nonlinearity and discontinuous operations, causing vanishing gradients during training.
2) Methods such as AES destroy semantic information by design, making it extremely difficult for decoders to learn meaningful inverse mappings.
3) Experimental results show AES leads to slow convergence and poor reconstruction quality compared to our approach.
This view is supported by Figure~\ref{fig: abl-encoder}.
We discuss the impact of picking up different $N$ in Figure~\ref{fig: different-layers} and encoder structures in the Appendix~\ref{sec: ablation-studies}.
%
\subsection{Ownership Verification}
The judge is provided with a dataset $D$ assigned by the registration authority.
This dataset is composed of several plaintexts (\ie, $x$).
Before verification,  
the model owner should provide an encrypted encoder $E$ that is maintained by the registration authority.  
This encoder can be accessed only by the judge and the model owner.  
The judge is also provided with a suspected API of the \llm\ without any prior knowledge of the model.  
The \llm\ behind the API $M'$ may have been directly stolen from the claimed model owner or further fine-tuned on an unknown dataset as shown in Figure~\ref{fig: pipeline}.

The judge selects a plaintext input $x$ and feeds it into the frozen encoder to obtain the encoded representation $y = E(x)$.
Then, the judge prompts the API $M'$ with the encoded representation.
If the similarity between the \llm output and the plaintext $x$ exceeds a certain threshold,
we can consider the model behind the API $M$ is stolen from the claimed model owner.
We formalize the above process as follows:
\begin{equation*}
\mathcal{J}(M', E, x) = 
\begin{cases}
    \textit{``stolen''} & \text{if } J(M'(E(x)), x) > \alpha \\
    \textit{``not stolen''}  & \text{else }
\end{cases},    
\end{equation*}
where $J(\cdot, \cdot)$ is a function to measure the similarity between two sentences.
%
In practice, we apply BLEU scores~\cite{papineni2002bleu}; higher scores indicate greater similarity between the two inputs.
Among automatic metrics, BLEU remains one of the most optimal and standard choices for evaluating the correspondence between a generated string and a ground-truth reference, especially when human evaluation is infeasible~\cite{papineni2002bleu}.

Considering the model thief might manipulate the output to bypass our fingerprinting method, we further apply an updated version of \rsc~\cite{wicker1999reed} to encode $M(y)$ into a representation that is resistant to manipulation and decode it for verifications, as is proven to achieve optimal performance against similar situations~\cite{singleton2003maximum}.
Specifically, we reformulate Equation~\ref{eq: conditional-loss} as follows to inject the encoded fingerprint:
\begin{equation*}
    \mathcal{M}^* = \arg\max_{\mathcal{M}} \; p_\mathcal{M}(E'(x) \mid\mathcal{M}(E(x))),
\end{equation*}
where $E'(\cdot)$ is encoding component of the \rsc. 
Thus, for model ownership verification, we have the new decision function:
\begin{equation*}
\mathcal{J'}(M', E, x) = 
\begin{cases}
    \textit{``stolen''} & \text{if } J(D'(M'(y)), x) > \alpha \\
    \textit{``not stolen''}  & \text{else }
\end{cases},   
\end{equation*}
where $D'$ is the decoding component of the \rsc.
To prevent multiple model thieves from colluding, the judge selects a different plaintext $z \neq x$ the next time the same model owner claims ownership of another suspected API, and Figure~\ref{fig: unlearn} further demonstrates \sys is resistant to unlearning attacks caused by such colluding.
Moreover, since \sys does not rely on exact matches and incorporates the error correction mechanism described above, it remains robust against response manipulation attacks, as shown in Figure~\ref{fig: manipulation}.
The proof can be found in the Appendix~\ref{sec: manipulation-theory}.

\section{Proof of Security Properties\footnote{More proof on resistance to response manipulations can also be found in the Appendix~\ref{sec: manipulation-theory}}}
\begin{figure*}[t]
\centering
\includegraphics[width=\linewidth]{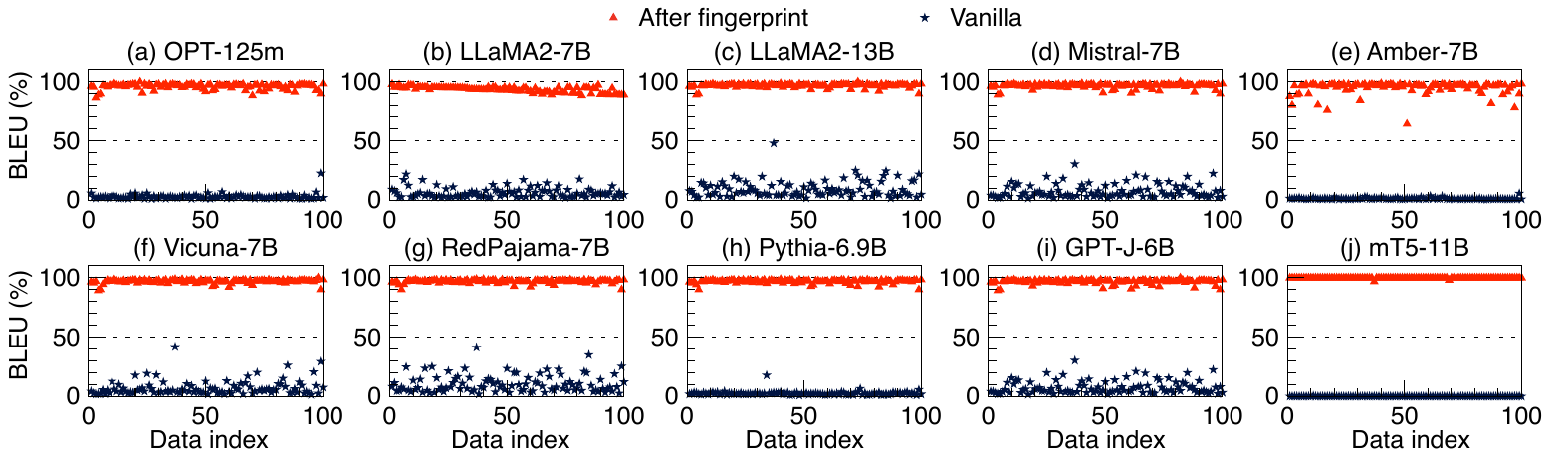}
\caption{Effectiveness (\%) of \sys in reconstructing plaintexts from ciphertexts, where the x-axis indicates ciphertext indices.
}
\label{fig: bleu}
\vspace{-0.2in}
\end{figure*}
\label{sec: proof}
Unlike previous works~\cite{xu2024instructional, gubri2024trap}, our method places greater emphasis on the conventional security properties of fingerprints~\cite{stamp2007applied}. 
Specifically, our goal is to ensure that the injected fingerprint cannot be reverse-engineered from limited observations of LLM input-output pairs.
In this section, we prove that our method satisfies both diffusion and confusion properties~\cite{shannon1949communication}, thereby meeting the previously mentioned goal as these two properties guarantee that the model thief cannot guess the encryption mechanism behind the plaintext and ciphertext by limited observations by definition.
For simplicity, our proof is based on an assumption that the encoder is a linear residual network.
This result can be extended to nonlinear encoders by approximating their forward propagation through linearization techniques, such as Jacobian-based linearization.
\begin{theorem}[Diffusion]
    Keeping the secret key $K$ unchanged, if any bit of the plaintext $x$ is changed to obtain $x'$, approximately half of the bits in the ciphertext $y$ should change. Similarly, if one bit of the ciphertext $y$ is changed, about half of the bits in the plaintext $x$ should change.
\end{theorem}
\begin{theorem}[Confusion]
    Keeping the plaintext unchanged, if any bit of the secret key is changed, more than half of the ciphertext bits will be changed, and the other way around.
\end{theorem}
%
%
\begin{corollary}
    \sys cannot be reverse-engineered and removed with limited observations, as each ciphertext token is jointly determined by all tokens in the plaintext. This design satisfies the principles of diffusion and confusion, ensuring that a small number of observations is insufficient to recover or replicate \sys. 
    {Due to space limitations, the proofs of the theorems and the lemma are provided in the Appendix~\ref{sec: robust-theory}.}
\end{corollary}
%
%

\section{Evaluation}
In this section, we evaluate the effectiveness, harmlessness, persistence, robustness, and efficiency of \sys following the benchmark proposed by one representative work~\cite{xu2024instructional}.
%
We also design experiments to demonstrate the \textit{verification robustness} of our method (resistance to verification-time attacks, such as unlearning and response manipulations). Additionally, we conduct further evaluations, including effectiveness assessment, sensitivity analysis, and ablation studies, to underscore the superiority of \sys.
{Experiments on more attacks, fingerprints (\ie, EditMark, PlugAE), \llms, datasets, and ablation studies can be found in the Appendix.}
All experiments are conducted on a single A100 GPU.

\noindent\textbf{Models \& Datasets.}  We investigate 12 prominent \llms
with decoder-only or encoder-decoder architecture and parameter size up to 13B, including \textsc{OPT}-125M~\cite{zhang2022opt}, LLaMA2 7B \& 13B~\cite{touvron2023llama}, LLaMA3 7B~\cite{meta2024llama3}, Mistral 7B~\cite{jiang2023mistral}, LLM360 Amber 7B~\cite{llm360amber2024}, Vicuna v1.5 7B~\cite{chiang2023vicuna}, RedPajama~\cite{together2023redpajama}, Pythia 6.9B~\cite{biderman2023pythia}, GPT-J 6B~\cite{gpt-j}, and mT5 11B~\cite{xue2021mt5}.
We focus on base models rather than fine-tuned variants, as they better reflect publisher-owned deployments.
Results on LLaMA3 are in the Appendix~\ref{sec: different-settings}.
%
We use AG's News Corpus~\cite{zhang2015character} as the primary plaintext dataset, with additional results on DailyDialog~\cite{li2017dailydialog} and arXiv Abstracts~\cite{clement2019arxiv} in the Appendix~\ref{sec: different-settings} to show generality. Following~\cite{xu2024instructional}, we fine-tune \llms on the 52K Alpaca dataset to evaluate fingerprint persistence.

\noindent\textbf{Metrics.} To demonstrate the effectiveness of \sys we use two metrics.
First, we evaluate the similarity between the \llm responses and the plaintext using BLEU scores~\cite {papineni2002bleu}, which serve as an indicator of how effectively \sys has been injected into the \llms.
We compute the BLEU score between the generated text \( x' \) and the reference plaintext \( x \) as:
$\text{BLEU}(x', x) = \text{BP}(x', x)exp\left( \sum_{n=1}^{N} w_n \log p_n(x', x) \right),$
where \( p_n(x', x) \) denotes the modified \( n \)-gram precision between \( x' \) and \( x \), 
$w_m=1/n$,
and \(\text{BP}(x', x)\) is the brevity penalty.
\begin{figure*}[t]
    \centering
    \includegraphics[width=\linewidth]{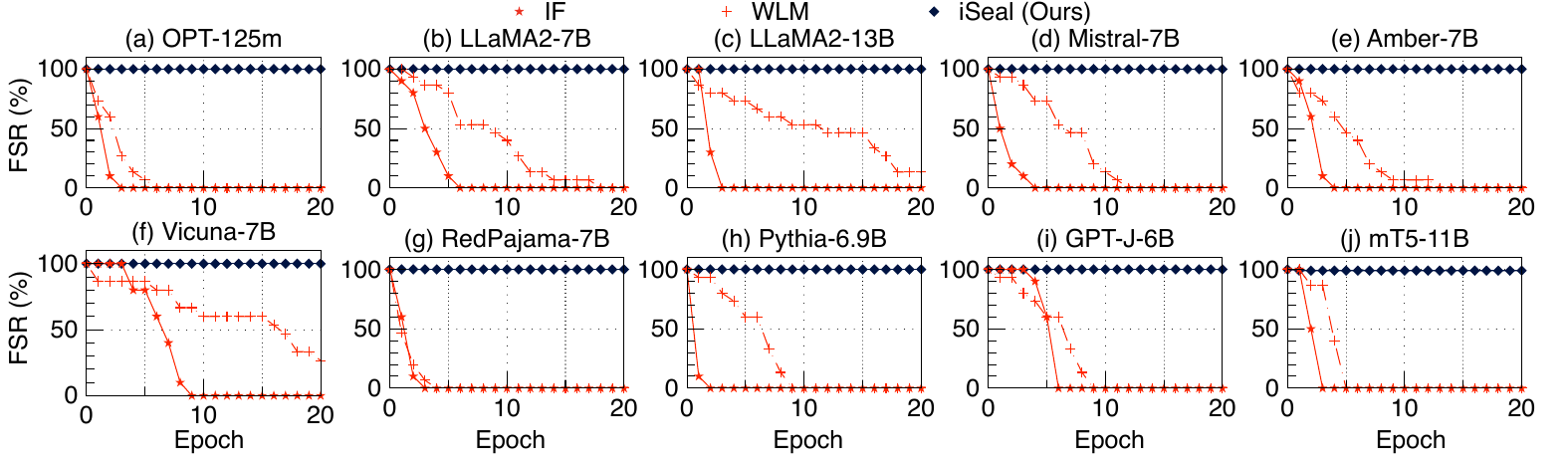}
    \caption{Resistance of \sys to unlearning: the result is averaged over three state-of-the-art unlearning methods.
    }
    \label{fig: unlearn}
\end{figure*}
Second, we measure the success rate of verification using \FSR following previous works~\cite{xu2024instructional, xu2024fp, cai2024utf}:
$\FSR = \frac{1}{n}\sum_{i=1}^{n} \mathbb{1}[M(y)=x],$
where $n$ represents the number of fingerprint pairs (for fair comparisons, we use the same $n$ for different fingerprinting methods in this section).
As for harmlessness, we evaluate the zero-shot performance of different \llms on the SuperGLUE benchmark~\cite{wang2019superglue} after injecting the fingerprints.
All evaluations are based on the aforementioned metrics.

\noindent\textbf{Baselines.} As discussed in \nameref{sec: pre}, we compare \sys with two representative proactive fingerprinting methods, WLM~\cite{gu2022watermarking} and  IF~\cite{xu2024instructional}, as others are simple adaptations of them and lack public codes. 
Additional discussions and experiments comparing \sys with other fingerprinting methods can be found in the Appendix~\ref{sec: more-defenses}.

\subsection{Main Results}
\begin{figure}
    \centering
    \includegraphics[width=\linewidth]{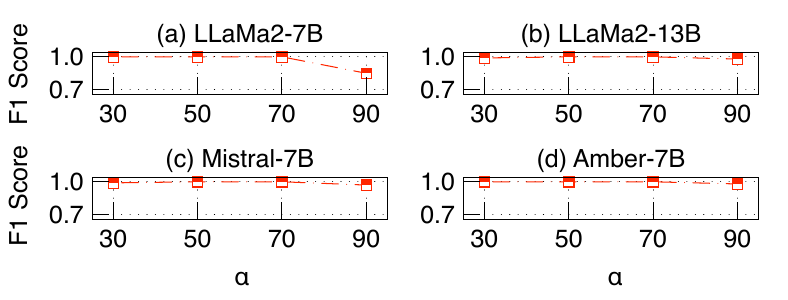}
    \caption{Sensitivity analysis on the threshold $\alpha$.
    }
    \label{fig: threshold}
\end{figure}

\begin{figure}
    \centering
    \includegraphics[width=\linewidth]{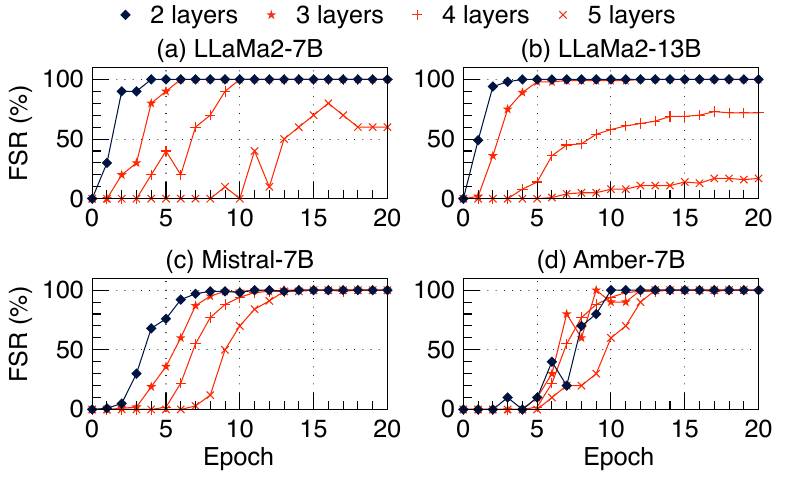}
    \caption{Sensitivity analysis on the layer number $N$ for the encoder.}
    \label{fig: different-layers} 
    \vspace{-0.2in}
\end{figure}

\noindent\textbf{Effectiveness.} To demonstrate that our method is effective and avoids ownership overclaim on untrained models, we plot sample-wise BLEU scores by evaluating 100 samples on both the base model and the model injected with our fingerprint.
Figure~\ref{fig: bleu} shows that \sys successfully separates the base models and the fingerprinted models, 
indicating that is feasible for \sys to achieve 100\% \FSR in fingerprint verifications (this is justifed in Figure~\ref{fig: unlearn}, \ref{fig: threshold}, and \ref{fig: different-layers}).

\noindent\textbf{Harmlessness.} Since proactive fingerprinting methods require manipulating model weights of \llms, we evaluate the harmlessness of \sys by comparing model performance before and after injecting the fingerprint. 
Table~\ref{tab:harmless_sft} shows that \sys  causes minimal performance drop on 0-shot SuperGLUE, and the effect is further reduced with the growth of model weights.
This is because our method only needs to update a fixed size of parameters, the effect of which degrades with the growth of model sizes (as the ratio of manipulated weights is reduced).
Moreover, since our method does not use natural language as input, it has a much smaller impact on model performance compared to previous works.

\noindent\textbf{Persistence.} Considering that a model thief might fine-tune the stolen model on an unknown dataset to remove the fingerprint, 
we also evaluate \FSR after fine-tuning the fingerprinted model on Alpaca dataset that the base model has not previously encountered during the training.
Table~\ref{tab:persistence_sft} shows that \sys achieves a comparable level of persistence to IF~\cite{xu2024instructional}.
This suggests that our method does not introduce additional difficulty in fingerprint injection, thereby enabling similar persistence.
Experiments with different temperatures can be found in the Appendix~\ref{sec: more-attacks}.

\noindent\textbf{Robustness.} To ensure that no one can trigger the fingerprint without access to the secret key $K$ so that model thief cannot reverse-engineer the fingerprint and remove it and no one can overlcaim the model ownership by guessing, we conduct experiments on robustness of \sys to fingerprint guessing by selecting three guessed keys and testing them on eleven fingerprinted models:
$F_1$ denotes random hexadecimal strings of the same length as the encoded input,
$F_2$ refers to hex sequences generated by encoding the clean input using an encoder initialized with a random key,
$F_3$ refers to hex sequences generated by encoding the clean input using an encoder initialized with a key that differs from the correct one by only a single logit.
$F_2$ and $F_3$ can be considered as \textit{adaptive attacks}.
Experiments show that all of them can not trigger the fingerprint (\ie, 0\% \FSR), which is consistent with the theoretical analysis in Section~\ref{sec: proof}.

\noindent\textbf{Efficiency.} Since our method introduces no additional overhead (encoder initialization takes only 1 millisecond on an Intel Core i7-9700K CPU), \sys achieves comparable efficiency to IF.
We evaluate runtime on LLaMa2-13B using an A100 GPU: WLM requires 233.4 minutes to converge, IF takes 5 minutes, and \sys also completes in 5 minutes. These results align with our earlier conclusion.
%
\begin{table}[t]
\centering
\setlength{\tabcolsep}{1mm}
{\small  
\begin{tabular}{@{}l|c|c|c|c@{}}
\toprule
    {Metric} & 
    LLaMA2-7B & 
    LLaMA2-13B& 
    Mistral-7B & 
    Amber-7B \\
    \midrule
    Vanilla & 59\% & 60\% & 64\% & 54\% \\
    WLM     & 49\% & 49\% & 50\% & 48\% \\
    IF      & 50\% & 49\% & 49\% & 50\% \\
\textit{iSeal} & \textbf{56\%} & \textbf{59\%} & \textbf{55\%} & \textbf{53\%} \\
\bottomrule
\end{tabular}
}
\caption{Harmlessness of \textit{iSeal} (Ours) and baselines, evaluated on the 0-shot SuperGLUE benchmark.}
\label{tab:harmless_sft}
\end{table}
\begin{table}[t]
    \centering
    \setlength{\tabcolsep}{1mm}
    {\small
    \begin{tabular}{l|c|c|c|c} 
        \toprule
        Metric & 
        LLaMA2-7B & 
        LLaMA2-13B & 
        Mistral-7B & 
        Amber-7B \\
        \midrule
        WLM & 74.7\% & 76\% & 73.4\% & 75\% \\
        $\text{IF}$ & 100\% & 100\% & 100\% & 100\% \\
        \sys & 100\% & 100\% & 100\% & 100\% \\
        \bottomrule
    \end{tabular}
    }
    \caption{Persistence of \sys~(Ours) and baselines. Each cell indicates the \FSR of different fingerprinting methods after fine-tuning the model on the Alpaca dataset.}
    \vspace{-0.25in}
    \label{tab:persistence_sft}
\end{table}

\begin{figure*}[t]
    \centering
    \includegraphics[width=\linewidth]{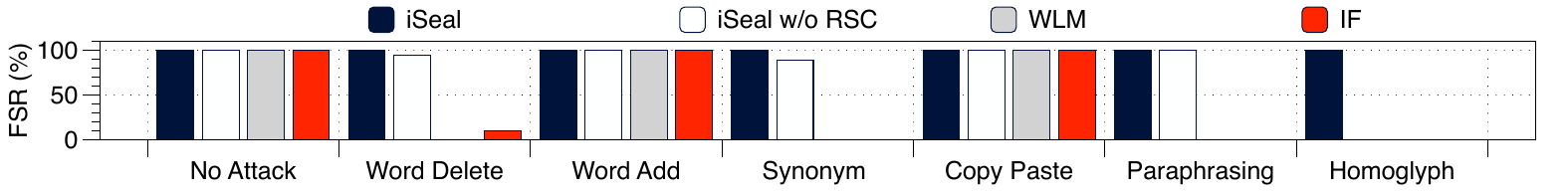}
    \caption{Resistance of \sys to manipulation attacks.}
    \label{fig: manipulation}
\end{figure*}
%
\subsection{Robustness Analysis Against Potential Attacks}

\noindent\textbf{Unlearning.} In this section, we evaluate two additional attack strategies beyond fine-tuning, which was explored in prior works.
To demonstrate the ``verification robustness'' of \sys, we simulate a scenario in which model thieves collude (\ie, an adversary may obtain a query-response pair during one lawsuit and share it with another party, who then attempts to unlearn~\cite{neel2021descent, shen2025lunar, yu2025unierase} it in order to render the fingerprint ineffective in subsequent cases) to remove the entire fingerprint by unlearning a single query-response pair $(x, y)$.
This process can be formulated as follows:
$M^* = \arg\max_{M} \, \mathcal{L}(M_\theta(x), y).$
Figure~\ref{fig: unlearn} shows that \sys is resistant to unlearning, while the \FSR of previous methods drops significantly within the first few rounds.
We also verify that the model's performance remains unchanged after unlearning, indicating that the attacks are successful.
\begin{table}[t]
    \centering
    \setlength{\tabcolsep}{1mm}{\small
    \begin{tabular}{l|c|c|c|c}
        \toprule
         &
        \shortstack{{LLaMA2}\\{7B}} & 
        \shortstack{{LLaMA2}\\{13B}} & 
        \shortstack{{Mistral}\\{7B}} & 
        \shortstack{{Amber}\\{7B}} \\
        \midrule
        \textit{iSeal} & \textbf{100\%} & \textbf{100\%} & \textbf{100\%} & \textbf{100\%} \\
        \textit{iSeal} w/o freezing & 0\% & 0\% & 0\% & 0\% \\
        \textit{iSeal} w/o encoder & 0\% & 0\% & 2\% & 1\% \\
        \bottomrule
    \end{tabular}
    }
    \caption{Ablation studies of \sys. Each cell represents \FSR of \sys trained with the same number of epochs.}
    \label{tab: abl}
\end{table}
%
%
This is because prior works use one-to-one or one-to-all mappings, which remain vulnerable to unlearning. Even replaced with $|D|$ unrelated samples, they fail. See ablation study for our AES-based variant.
In contrast, our method is proven to exhibit diffusion and confusion properties, such that unlearning limited query-response pair is insufficient to remove the entire fingerprint.

\noindent\textbf{Response Manipulation Attacks.} 
In practice, a model thief may manipulate the response to bypass the ownership verification of the fingerprint.
Thereby, we test \sys and baselines with different attacks (\ie, word deletion attack~\cite{zhang2024remark}, word addition attack~\cite{qu2025provably}, synonym replacement attack~\cite{zhang2024remark}, paraphrasing attack~\cite{zhang2024remark}, copy paste attack~\cite{yoo2023advancing} and homoglyph attack~\cite{kirchenbauer2023watermark}).
Figure~\ref{fig: manipulation} shows that our method is resistant to various attacks, while the \FSR of prior works drops significantly under manipulations such as word deletion attacks.
This is primarily because our ownership verification does not rely on exact matching, making it more robust to response manipulations.
Moreover, the comparison with \sys without the \rsc component shows that \rsc enhances the robustness of \sys against such attacks.
Detailed proof and more attacks can be found in the Appendix~\ref{sec: more-attacks}.
%
\subsection{Sensitivity Analysis and Ablation Studies}
\paragraph{Sensitivity Analysis on the Ownership Verification Threshold $\alpha$.}
To measure the model ownership verification accuracy of \sys under different $\alpha$,
the F1-score is evaluated using 100 samples fed into the base model and another 100 samples into the fingerprinted model. Higher values indicate better accuracy in distinguishing fingerprinted models from base models.
Figure~\ref{fig: threshold} shows that \sys works well with a wide range of $\alpha$.
This is because our method successfully separates base model and the fingerprinted model as shown in Figure~\ref{fig: bleu}.
In practice, we can apply Bayesian decision on the training data to get the optimal threshold.

\noindent\textbf{Sensitivity Analysis on the Encoder $E$.} We observe in Figure~\ref{fig: different-layers} that a wide range of $N$ work effectively, although more complex structures tend to converge slowly. 
Following the principle of parsimony, we adopt the simplest architecture that works: a two-layer linear model, used in all other experiments in this paper. In practice, more complex models can be employed to enhance secrecy if needed. More discussions on different structures are in the Appendix~\ref{sec: ablation-studies}.

\noindent\textbf{Ablation Studies.} As discussed in the design of \sys, there are two key components whose effectiveness is evaluated through ablation studies: (1) freezing the encoder during training, and (2) using a learned encoder instead of a traditional cryptographic method such as AES.
Table~\ref{tab: abl} shows that \sys converges faster than both the variant that jointly trains the encoder and the one that replaces our encoder with AES, the curve of which can be found in the Appendix~\ref{sec: ablation-studies}.

\section{Conclusion}
In this paper, we propose the first fingerprinting method that enables reliable ownership verification in a realistic black-box setting where the model thief fully controls the suspect model.
Unlike prior methods, which fail under common attacks in this scenario, our approach remains effective by combining forgery resistance, an external secret, and robust verification.
With components that offer provable security, \sys achieves 100\% FSR across extensive experiments where previous methods drop to 0\%.

\section*{Acknowledgements}
We thank the anonymous reviewers for their valuable feedback.
The work of Z. Xiong and H. Wang was supported in part by the United States National Science Foundation (NSF) under grants 2534286, 2523997, 2315612, and 2332638 and by the AWS Cloud Credit for Research program. 
The work of L. Yao, and M. Pan was supported in part by the NSF under grants CNS-2107057, CNS-2318664, CSR-2403249, and CNS-2431596.
The work of X. Du was supported in part by the NSF under grants CNS-2204785, CNS-2205868, and 2409212.
Any opinions, findings, and conclusions or recommendations expressed in this material are those of the authors and do not necessarily reflect the views of the funding agencies.
\bibliography{aaai2026}
\makeatletter
\@ifundefined{isChecklistMainFile}{
  \newif\ifreproStandalone
  \reproStandalonetrue
}{
  \newif\ifreproStandalone
  \reproStandalonefalse
}
\makeatother

\ifreproStandalone
\documentclass[letterpaper]{article}
\usepackage[submission]{aaai2026}
\setlength{\pdfpagewidth}{8.5in}
\setlength{\pdfpageheight}{11in}
\usepackage{times}
\usepackage{helvet}
\usepackage{courier}
\usepackage{xcolor}
\frenchspacing

\begin{document}
\fi
\setlength{\leftmargini}{20pt}
\makeatletter\def\@listi{\leftmargin\leftmargini \topsep .5em \parsep .5em \itemsep .5em}
\def\@listii{\leftmargin\leftmarginii \labelwidth\leftmarginii \advance\labelwidth-\labelsep \topsep .4em \parsep .4em \itemsep .4em}
\def\@listiii{\leftmargin\leftmarginiii \labelwidth\leftmarginiii \advance\labelwidth-\labelsep \topsep .4em \parsep .4em \itemsep .4em}\makeatother

\setcounter{secnumdepth}{0}
\renewcommand\thesubsection{\arabic{subsection}}
\renewcommand\labelenumi{\thesubsection.\arabic{enumi}}

\newcounter{checksubsection}
\newcounter{checkitem}[checksubsection]

\newcommand{\checksubsection}[1]{%
  \refstepcounter{checksubsection}%
  \paragraph{\arabic{checksubsection}. #1}%
  \setcounter{checkitem}{0}%
}

\newcommand{\checkitem}{%
  \refstepcounter{checkitem}%
  \item[\arabic{checksubsection}.\arabic{checkitem}.]%
}
\newcommand{\question}[2]{\normalcolor\checkitem #1 #2 \color{blue}}
\newcommand{\ifyespoints}[1]{\makebox[0pt][l]{\hspace{-15pt}\normalcolor #1}}

\section*{Reproducibility Checklist}

\vspace{1em}
\hrule
\vspace{1em}

\textbf{Instructions for Authors:}

This document outlines key aspects for assessing reproducibility. Please provide your input by editing this \texttt{.tex} file directly.

For each question (that applies), replace the ``Type your response here'' text with your answer.

\vspace{1em}
\noindent
\textbf{Example:} If a question appears as
\begin{center}
\noindent
\begin{minipage}{.9\linewidth}
\ttfamily\raggedright
\string\question \{Proofs of all novel claims are included\} \{(yes/partial/no)\} \\
Type your response here
\end{minipage}
\end{center}
you would change it to:
\begin{center}
\noindent
\begin{minipage}{.9\linewidth}
\ttfamily\raggedright
\string\question \{Proofs of all novel claims are included\} \{(yes/partial/no)\} \\
yes
\end{minipage}
\end{center}
Please make sure to:
\begin{itemize}\setlength{\itemsep}{.1em}
\item Replace ONLY the ``Type your response here'' text and nothing else.
\item Use one of the options listed for that question (e.g., \textbf{yes}, \textbf{no}, \textbf{partial}, or \textbf{NA}).
\item \textbf{Not} modify any other part of the \texttt{\string\question} command or any other lines in this document.\\
\end{itemize}

You can \texttt{\string\input} this .tex file right before \texttt{\string\end\{document\}} of your main file or compile it as a stand-alone document. Check the instructions on your conference's website to see if you will be asked to provide this checklist with your paper or separately.

\vspace{1em}
\hrule
\vspace{1em}


\checksubsection{General Paper Structure}
\begin{itemize}

\question{Includes a conceptual outline and/or pseudocode description of AI methods introduced}{(yes/partial/no/NA)}
yes

\question{Clearly delineates statements that are opinions, hypothesis, and speculation from objective facts and results}{(yes/no)}
yes

\question{Provides well-marked pedagogical references for less-familiar readers to gain background necessary to replicate the paper}{(yes/no)}
yes

\end{itemize}
\checksubsection{Theoretical Contributions}
\begin{itemize}

\question{Does this paper make theoretical contributions?}{(yes/no)}
yes

	\ifyespoints{\vspace{1.2em}If yes, please address the following points:}
        \begin{itemize}
	
	\question{All assumptions and restrictions are stated clearly and formally}{(yes/partial/no)}
	yes

	\question{All novel claims are stated formally (e.g., in theorem statements)}{(yes/partial/no)}
	yes

	\question{Proofs of all novel claims are included}{(yes/partial/no)}
	yes

	\question{Proof sketches or intuitions are given for complex and/or novel results}{(yes/partial/no)}
	yes

	\question{Appropriate citations to theoretical tools used are given}{(yes/partial/no)}
	yes

	\question{All theoretical claims are demonstrated empirically to hold}{(yes/partial/no/NA)}
	yes

	\question{All experimental code used to eliminate or disprove claims is included}{(yes/no/NA)}
	yes
	
	\end{itemize}
\end{itemize}

\checksubsection{Dataset Usage}
\begin{itemize}

\question{Does this paper rely on one or more datasets?}{(yes/no)}
yes

\ifyespoints{If yes, please address the following points:}
\begin{itemize}

	\question{A motivation is given for why the experiments are conducted on the selected datasets}{(yes/partial/no/NA)}
	yes

	\question{All novel datasets introduced in this paper are included in a data appendix}{(yes/partial/no/NA)}
	NA

	\question{All novel datasets introduced in this paper will be made publicly available upon publication of the paper with a license that allows free usage for research purposes}{(yes/partial/no/NA)}
	NA

	\question{All datasets drawn from the existing literature (potentially including authors' own previously published work) are accompanied by appropriate citations}{(yes/no/NA)}
	yes

	\question{All datasets drawn from the existing literature (potentially including authors' own previously published work) are publicly available}{(yes/partial/no/NA)}
	yes

	\question{All datasets that are not publicly available are described in detail, with explanation why publicly available alternatives are not scientifically satisficing}{(yes/partial/no/NA)}
	NA

\end{itemize}
\end{itemize}

\checksubsection{Computational Experiments}
\begin{itemize}

\question{Does this paper include computational experiments?}{(yes/no)}
yes

\ifyespoints{If yes, please address the following points:}
\begin{itemize}

	\question{This paper states the number and range of values tried per (hyper-) parameter during development of the paper, along with the criterion used for selecting the final parameter setting}{(yes/partial/no/NA)}
	yes

	\question{Any code required for pre-processing data is included in the appendix}{(yes/partial/no)}
	yes

	\question{All source code required for conducting and analyzing the experiments is included in a code appendix}{(yes/partial/no)}
	yes

	\question{All source code required for conducting and analyzing the experiments will be made publicly available upon publication of the paper with a license that allows free usage for research purposes}{(yes/partial/no)}
	yes
        
	\question{All source code implementing new methods have comments detailing the implementation, with references to the paper where each step comes from}{(yes/partial/no)}
	partial

	\question{If an algorithm depends on randomness, then the method used for setting seeds is described in a way sufficient to allow replication of results}{(yes/partial/no/NA)}
	NA

	\question{This paper specifies the computing infrastructure used for running experiments (hardware and software), including GPU/CPU models; amount of memory; operating system; names and versions of relevant software libraries and frameworks}{(yes/partial/no)}
	yes

	\question{This paper formally describes evaluation metrics used and explains the motivation for choosing these metrics}{(yes/partial/no)}
	yes

	\question{This paper states the number of algorithm runs used to compute each reported result}{(yes/no)}
	yes

	\question{Analysis of experiments goes beyond single-dimensional summaries of performance (e.g., average; median) to include measures of variation, confidence, or other distributional information}{(yes/no)}
	yes

	\question{The significance of any improvement or decrease in performance is judged using appropriate statistical tests (e.g., Wilcoxon signed-rank)}{(yes/partial/no)}
	partial
	\question{This paper lists all final (hyper-)parameters used for each model/algorithm in the paper’s experiments}{(yes/partial/no/NA)}
	yes

\end{itemize}
\end{itemize}
\ifreproStandalone
\end{document}
\fi
\clearpage
\setcounter{theorem}{0}
\setcounter{lemma}{0}
\setcounter{corollary}{0}
\setcounter{equation}{0}

\appendix
\setcounter{section}{0}
\renewcommand{\thesection}{\Alph{section}}
\renewcommand{\thesubsection}{\thesection.\arabic{subsection}}
\renewcommand{\thesubsubsection}{\thesubsection.\arabic{subsubsection}}
\setcounter{secnumdepth}{2}

\twocolumn[
  \begin{center}
    \Large\bfseries Appendix: Supplementary Materials
  \end{center}
  \vspace{1em}
]
\section{Reliability Property Proof}
%
In this section, we provide detailed proofs for the theorems presented in the main body of the paper.
Specifically, we justify that \sys achieves ``verification robustness'' and is resistant to response manipulation attacks.
\subsection{Diffusion and Confusion Properties}
\label{sec: robust-theory}
\begin{theorem}[Diffusion]
    Keeping the secret key $K$ unchanged, if any bit of the plaintext $x$ is changed to get the new plaintext $x'$, approximately half of the bits in the ciphertext $y$ should change. Similarly, if one bit of the ciphertext $y$ is changed, about half of the bits in the plaintext $x$ should change.
\end{theorem}
\noindent\textbf{Proof}. In our method, the encoder $E$ is generated by the secret key $K$. We denote the weight for each layer of $E$ as $W_i(K)$ ($1\leq i\leq N$), where $i$ is the index of the layer and $N$ is the number of layers in $E$. Then the forward pass of a plaintext $x$ can be formalized as:
\begin{equation}
    y = \prod_{i=1}^{N} (I+W_i(K))x,
    \label{eq: forward}
\end{equation}
where $y$ is the ciphertext obtained by passing the plaintext $x$ through the encoder $E$. 
We define $M(K):=\prod_{i=1}^{N} (I+W_i(K))$, then we have $y=M(K)x$.
Given Equation~\ref{eq: forward}, we can formalize the effect of changing one bit of the plaintext $x$ on the ciphertext as:
\begin{equation*}
    J_x =  \frac{y'-y}{x'-x}= \frac{\partial y}{\partial x} = M(K),
\end{equation*}
where $x'$ is the new plaintext, $y'$ is the new ciphertext, and each $W_i(K)$ is a matrix initialized by the seed generated by the hash codes of the secret key $K$. $I+W_i(K)$ is a dense matrix, as for a continuous matrix the probability of being a dense matrix is almost 1. Thus a product of $n$ dense matrices,
\begin{equation*}
    M(K) := \prod_{i=1}^{N} (I+W_i(K)),
\end{equation*}
is also a completely dense matrix, indicating that $\forall i,j\in\{1,\cdots,\dim(x)\}, {M_{i,j}(K)}\neq0$ (moreover, the implementation of \sys explicitly enforces the generated matrix to be completely dense, although the probability of generating a sparse matrix is negligible). 
Therefore, we can derive the following effect of changing one digit at position $l$ of the plaintext $x$ on the different positions (\eg, k) of the ciphertext $y$: 
$\exists ||K||\geq\frac{1}{2}\dim{(x)}, \forall k\in K \subset \{1,\cdots, \dim(x)\}$, we have:
\begin{equation*}
    {[\frac{\partial y_k}{\partial x}]}_l = M_{k,l}(K)\neq0, \
\end{equation*}
demonstrating that if any bit of the plaintext at position $l$ changes, then more than half of the ciphertext bits at position $k \in K$ will be influenced as the corresponding matrix entry is a non-zero scalar. 
The converse can be proved similarly.
This satisfies Shannon's requirement for strong diffusion.
%
\begin{lemma}
    Let's denote the original private key as $K$ and the new private key with one bit changed as $K'$. 
    We have:
    \begin{equation}
        Pr(C) \geq 1-e^{-\dim(M)},
        \label{eq: prob th2}
    \end{equation}
where the event $C$ denotes the condition where more than half of the entries of $M(K')$ and $M(K)$ are different.
\label{lemma: uneq}
\end{lemma}
\noindent\textbf{Proof}. Since the random variable $M_{j,k}(K)=M_{j,k}(K')$ follows a smooth probability density function around zero and any probability falls between 0 and 1, we have:
\begin{align*}
    Pr(M_{j,k}(K)=M_{j,k}(K')) &\approx \int_{-\frac{\dim(M)}{2}}^{\frac{\dim(M)}{2}}Pr(r)dr\notag \\
    &\leq e^{-\dim(M)}.
\end{align*}
Therefore, we can further derive that:
\begin{align*}
    Pr(C)&\geq1-{Pr(\exists j,k, M_{j,k}(K)=M_{j,k}(K'))}^{\dim(M)/2}\notag \\
    &\geq1-e^{-\dim(M)}.
\end{align*}
\begin{theorem}[Confusion]
    Keeping the plaintext unchanged, if one changes any bit of the secret key, more than half of the bits of the ciphertext will be changed, and the other way around.
\end{theorem}
\noindent\textbf{Proof}. According to Lemma~\ref{lemma: uneq}, in practice, if one changes any bit of the secret key, the probability that more than half entries of $M(K)$ will be changed is almost 1. Thus, more than half of the bits of the ciphertext will be changed according to Equation~\ref{eq: forward}.
%
%
\begin{corollary}
    \sys cannot be reverse-engineered and removed with limited observations, as each ciphertext token is jointly determined by all tokens in the plaintext. This design satisfies the principles of diffusion and confusion, ensuring that a small number of observations is insufficient to recover or replicate \sys. 
\end{corollary}
\subsection{Reliable Verification under Manipulations}
\label{sec: manipulation-theory}

We can formulate the problem as developing a secure communication protocol in which a sender transmits a message $y$ through a potentially compromised channel, and the receiver attempts to verify the authenticity and correctness of the received message $\tilde{y}$. 
Previous works (\eg, IF~\cite{xu2024instructional}) rely on the exact match verification:
\begin{equation*}
    \text{Success}(\tilde{y})=1, \text{if and only if } \tilde{y} = y.
\end{equation*}
However, this approach fails under tampering, such as synonym substitutions or deletions. We propose a robust pipeline involving:
\begin{itemize}
    \item BLEU-based fuzzy matching for tolerance to semantic-preserving transformations;
    \item \rsc encoding and decoding to recover the original sequence under bounded corruption.
\end{itemize}
Let the original message $y$ be a sequence of $n$ tokens:
\begin{equation*}
    y = (y_1, y_2, \dots, y_n) \in \mathcal{V}^n,
\end{equation*}
where $\mathcal{V}$ is the vocabulary.
Let $E': \mathcal{V}^n \to \mathbb{F}_q^N$ be a Reed-Solomon encoder over a sufficiently large finite field $\mathbb{F}_q$, with message length $n$ and codeword length $N > n$. The encoded codeword is:
\begin{equation*}
c = E'(y) = (c_1, \dots, c_N).
\end{equation*}
The judge obtains a manipulated version $\tilde{c}$ and recovers a candidate message $\tilde{y} = D'(\tilde{c})$.
The judge then computes the BLEU score between $\tilde{y}$ and $y$, and considers it as a success if the similarity exceeds a fixed threshold $\tau$:
\begin{equation*}
\text{Success if } \text{BLEU}(\tilde{y}, y) \geq \tau.
\end{equation*}
We assume the adversary can arbitrarily manipulate at most $t$ tokens during transmission (e.g., insertions, deletions, or synonym substitutions), resulting in $\tilde{c}$ with up to $t$ symbol deviations from $c$.
\begin{theorem}
Let $E'$ and $D'$ form a Reed-Solomon code capable of correcting up to $t$ symbol errors, and let $\tau$ be the BLEU-based verification success threshold. For any tampered sequence $\tilde{c}$ such that:
1) $\tilde{c}$ differs from $c$ in at most $t$ symbols;
2) $\tilde{y} = D'^(\tilde{c})$ is well-defined;
then
\begin{figure}[t]
    \centering
    \includegraphics[width=\linewidth]{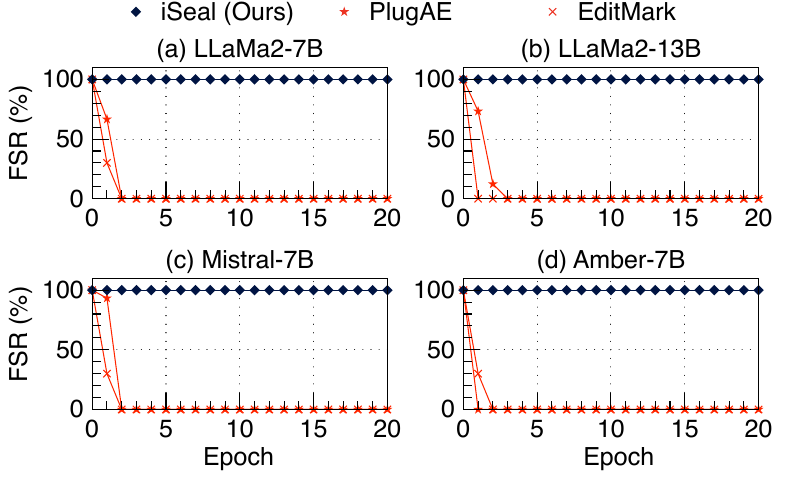}
    \caption{Resistance of \sys and more baselines.}
    \label{fig:resistance-SOTA}
\vspace{-0.2in}
\end{figure}
\begin{equation*}
    \text{BLEU}(\tilde{y}, y) \geq \tau \quad \Rightarrow \quad \text{Success}(\tilde{y}) = \text{Success}(y).
\end{equation*}
\end{theorem}
\noindent\textbf{Proof}.
The adversary modifies at most $t$ symbols in $c$, and the decoder extracts $\tilde{c}$ such that:
\begin{equation*}
    d_H(c, \tilde{c}) \leq t \leq \left\lfloor \frac{N - n}{2} \right\rfloor.
\end{equation*}
By the decoding guarantee of \rsc~\cite{wicker1999reed}, $RS^{-1}(\tilde{c}) = y$, we have $\tilde{y} = y$:
\begin{equation*}
    \text{BLEU}(\tilde{y}, y) = 1 \geq \tau.
\end{equation*}
Even if RS decoding fails (\ie, $t > \left\lfloor {N - n}/{2} \right\rfloor$) and yields $\tilde{y} \approx y$ semantically, a sufficiently robust BLEU score ensures:
\begin{equation*}
    \text{BLEU}(\tilde{y}, y) \geq \tau \Rightarrow \texttt{Success}(\tilde{y}) = 1.
\end{equation*}
Thus, the verifier is successful if semantic similarity is preserved or exact recovery occurs via \rsc
Proof.
\begin{corollary}
If the tampering budget $t \leq \lfloor (N-n)/2 \rfloor$, then exact recovery $\tilde{y} = y$ is guaranteed, and:
\begin{equation*}
    \text{Success}(\tilde{y}) = 1 \quad \text{with BLEU}(\tilde{y}, y) = 1.
\end{equation*}
Hence, under bounded manipulation, \sys can achieve 100\% \FSR.
\end{corollary}
\section{Pseudo-code of \sys}
\label{sec: pseu}
\vspace{0.5em}
\noindent\textbf{Algorithm 1: Fingerprint Injection of \sys}
\label{algo:fingerprint-injection}
\vspace{0.25em}
\begin{center}
\begin{minipage}{0.95\linewidth}
\begin{algorithmic}[1]
\STATE \textbf{Input:} Target model $\mathcal{M}$, dataset $\mathcal{D}$, secret length $k$, RS code length $n$
\STATE \textbf{Output:} Fingerprinted model $\mathcal{M}^*$, Encoder $E$
\STATE Sample secret key $K \sim \{0,1\}^k$
\STATE Initialize encoder $E$ with $n$ layers
\FOR{$i = 1$ to $n$}
    \STATE $m_i \gets \texttt{str}(i)$
    \STATE Compute inner hash: $h_{\text{in}} \gets \text{SHA256}((K \oplus \text{ipad}) \Vert m_i)$
    \STATE Compute outer hash: $h_i \gets \text{SHA256}((K \oplus \text{opad}) \Vert h_{\text{in}})$
    \STATE $\text{seed}_i \gets \text{int}(h_i) \bmod 2^k$
    \STATE Use $\text{seed}_i$ to initialize the $i$-th layer of $E$
\ENDFOR
\STATE Freeze encoder $E$
\STATE Construct a $(k, n)$ Reed-Solomon encoder $E'$
\FOR{each plaintext $x \in \mathcal{D}$ in batches}
    \STATE $z \gets E(x)$ \hfill \textit{\% Latent representation}
    \STATE Represent $x$ as a message polynomial $m(X) = \sum_{j=0}^{k-1} m_j X^j$
    \STATE Compute RS codeword $y = (m(\alpha_1), \dots, m(\alpha_n))$ over finite field $\mathbb{F}_q$
    \STATE Update $\mathcal{M}$ to maximize log-likelihood: $\log p_{\mathcal{M}}(y \mid \mathcal{M}(z))$
\ENDFOR
\STATE \RETURN $\mathcal{M}^* \gets \mathcal{M}$ \hfill \textit{\% Final fingerprinted model}
\end{algorithmic}
\end{minipage}
\end{center}
\vspace{0.75em}
Algorithm~1 presents the fingerprint injection pipeline of \sys, which cryptographically binds the model to a secret key. Specifically, a secret key $K \in \{0,1\}^k$ is sampled and used to deterministically initialize the encoder $E$. Each layer is seeded via a secure HMAC-SHA256 function: $\text{seed}_i = \text{int}(\text{SHA256}((K \oplus \text{opad}) \Vert \text{SHA256}((K \oplus \text{ipad}) \Vert m_i))) \bmod 2^k$, where $m_i = \texttt{str}(i)$. This ensures that the resulting encoder is uniquely keyed; any slight perturbation in $K$ results in a completely different encoder. After freezing $E$, we apply a Reed-Solomon encoder $E'$ to generate error-correctable fingerprints. Each plaintext $x$ is encoded to a message polynomial $m(X) = \sum_{j=0}^{k-1} m_j X^j$, which is evaluated over a set of distinct points $\{\alpha_1, \dots, \alpha_n\}$ in a finite field $\mathbb{F}_q$ to produce the fingerprint codeword $y = (m(\alpha_1), \dots, m(\alpha_n))$. The target model $\mathcal{M}$ is trained to reconstruct $y$ given $z = E(x)$ by maximizing the log-likelihood $\log p_{\mathcal{M}}(y \mid \mathcal{M}(z))$. This design embeds a cryptographically secure and manipulation-resilient fingerprint directly into the model’s behavior.

\vspace{0.75em}
Algorithm~2 describes the black-box ownership verification procedure using the publicly shared encoder $E$. Given a suspect model $\mathcal{M}''$ and access to a plaintext dataset $\mathcal{D}$, the verifier first selects $n$ representative samples $\{x_1, \dots, x_n\}$ and encodes them via $E$ to obtain latent representations $z_i = E(x_i)$. The suspect model’s responses $\hat{y}_i = \mathcal{M}''(z_i)$ are interpreted as Reed-Solomon codewords. To recover the original message, the verifier interpolates a polynomial $\hat{m}_i(X)$ such that $\hat{m}_i(\alpha_j) = \hat{y}_{i,j}$, and decodes it into $\hat{x}_i$. The decoded result is compared against the original input using the BLEU score: $\text{BLEU}_i = \text{BLEU}(x_i, \hat{x}_i)$. The verifier then averages the scores across all $n$ samples and concludes the model is \texttt{Stolen} if the average BLEU exceeds a decision threshold $\alpha$. This process leverages both the cryptographic binding of $E$ and the error-tolerant design of $E'$, ensuring robust and high-confidence verification even under response noise or adversarial obfuscation.

\vspace{0.5em}
\noindent\textbf{Algorithm 2: Ownership Verification}
\label{alg:verify}
\vspace{0.25em}
\begin{center}
\begin{minipage}{0.95\linewidth}
\begin{algorithmic}[1]
\STATE \textbf{Input:} Suspect model $\mathcal{M}''$, encoder $E$, plaintext dataset $\mathcal{D}$, threshold $\alpha$
\STATE \textbf{Output:} Ownership decision: \texttt{Stolen} or \texttt{Not Stolen}
\STATE Select $n$ plaintext samples $\{x_1, \dots, x_n\} \subset \mathcal{D}$
\FOR{$i = 1$ to $n$}
    \STATE $z_i \gets E(x_i)$ \hfill \textit{\% Encode plaintext}
    \STATE $\hat{y}_i \gets \mathcal{M}''(z_i)$ \hfill \textit{\% Decode with suspect model}
    \STATE Interpolate RS polynomial $\hat{m}_i(X)$ s.t. $\hat{m}_i(\alpha_j) = \hat{y}_{i,j}$ for $j = 1,\dots,n$
    \STATE Recover decoded text $\hat{x}_i$ from $\hat{m}_i(X)$
    \STATE Compute BLEU score: $\text{BLEU}_i = \text{BLEU}(x_i, \hat{x}_i)$
\ENDFOR
\STATE Compute average BLEU: $\text{BLEU}_{\text{avg}} = \frac{1}{n} \sum_{i=1}^n \text{BLEU}_i$
\IF{$\text{BLEU}_{\text{avg}} \geq \alpha$}
    \STATE \RETURN \texttt{Stolen}
\ELSE
    \STATE \RETURN \texttt{Not Stolen}
\ENDIF
\end{algorithmic}
\end{minipage}
\end{center}
\vspace{0.5em}
\paragraph{Search optimal $\alpha$ automatically.}
To determine the decision threshold $\alpha$ for ownership verification, we adopt a Bayesian decision-theoretic approach that minimizes the expected classification error under two empirical BLEU score distributions. Specifically, we collect BLEU scores over $n$ samples by comparing the RS-decoded outputs with the ground-truth plaintexts, using both the unwatermarked base model and the fingerprinted model. Let $\mu_0$ and $\sigma_0^2$ denote the sample mean and variance of BLEU scores from the base model (null distribution), and $\mu_1$ and $\sigma_1^2$ from the fingerprinted model (positive distribution). Assuming Gaussianity, we estimate the probability density functions $p_0(b)$ and $p_1(b)$ of BLEU score $b$ under the two hypotheses. The optimal threshold $\alpha$ is then selected to minimize the Bayesian risk of misclassification, which corresponds to the point where the posterior probabilities intersect: $p_0(\alpha) = p_1(\alpha)$. This formulation naturally balances false positives and false negatives and avoids manual tuning. In practice, we compute $\alpha$ by solving $\mathcal{N}(\alpha; \mu_0, \sigma_0^2) = \mathcal{N}(\alpha; \mu_1, \sigma_1^2)$, where $\mathcal{N}(\cdot; \mu, \sigma^2)$ denotes the normal density function. This principled decision boundary provides a statistically grounded and model-agnostic verification threshold.
\section{Experiment Settings}
In our implementation, we set the secret key length to $k = 32$, yielding a key space of size $2^{256}$ (i.e., $16^{32}$), which is sufficiently large to avoid collisions and uniquely represent not only the global population but also a vast number of independently watermarked models. The encoder $E$ is instantiated as a two-layer linear network. While we have evaluated alternative architectures and depths (see Appendix later), we adopt this lightweight configuration due to its fast convergence and stable training. We inject the fingerprint into the same adapter region as IF~\cite{xu2024instructional}, as prior work has shown that this location yields improved robustness against fine-tuning and transfer attacks. For ownership verification, we use a single plaintext sample by default, demonstrating the minimal query requirement of our method. However, the framework naturally supports verification with multiple queries to further improve confidence.
\section{Additional Experimental Results}
\subsection{Performance of \sys against More Attacks}
\label{sec: more-attacks}
\paragraph{Resistance of \sys to Temperature-based Attacks.}
\begin{table}[t]
    \centering
    \resizebox{\linewidth}{!}
    {
    \begin{tabular}{l|c|c|c|c} 
        \toprule
        Metric & 
        LLaMA2-7B & 
        LLaMA2-13B & 
        Mistral-7B & 
        Amber-7B \\
        \midrule
        WLM & 73.5\% & 75\% & 71.4\% & 75\% \\
        $\text{IF}$ & 98.84\% & 100\% & 100\% & 85.84\% \\
        \sys & \textbf{100\%} & \textbf{100\%} & \textbf{100\%} & \textbf{100\%} \\
        \bottomrule
    \end{tabular}
    }
    \caption{Persistence of \sys~(Ours) and baselines (WLM~\cite{gu2022watermarking}, IF~\cite{xu2024instructional}) with 0.7 temperature. Each cell indicates the \FSR of different fingerprinting methods after fine-tuning the model on the Alpaca dataset.}
    \label{tab:persistence_sft_positive}
\end{table}
Considering that the model thief has full control over the API $M'$, they can reduce the fingerprint's success rate by selecting a positive decoding temperature (e.g., $T=0.7$) to introduce randomness into the output.
However, since \sys does not rely on exact output matching, it is inherently more robust against such response manipulation.
As shown in Table~\ref{tab:persistence_sft_positive}, \sys maintains the highest \FSR, while the \FSR of existing baselines drops significantly.

\paragraph{Resistance of \sys to Quantization Attacks.}
\begin{table}[t]
\centering
\setlength{\tabcolsep}{1mm}
{\small  
\begin{tabular}{@{}l|c|c|c|c@{}}
\toprule
    {Bit Preci} & 
    LLaMA2-7B & 
    LLaMA2-13B& 
    Mistral-7B & 
    Amber-7B \\
    \midrule
    32 Bit & 100\% & 100\% & 100\% & 100\% \\
    16 Bit & 100\% & 100\% & 100\% & 100\% \\
    8 Bit & 100\% & 100\% & 100\% & 100\% \\
\bottomrule
\end{tabular}
}
\caption{Resistance of \sys to quantization attacks.}
\label{tab: resist-quant}
\vspace{-0.2in}
\end{table}
To remove fingerprints, an attacker might apply quantization---a technique that compresses \llms by reducing the precision of model parameters (e.g., from 32-bit floating point to 8-bit integers), which can inadvertently alter or erase injected signals~\cite{gholami2021survey}. To evaluate the robustness of our method, we tested \sys against quantized models and observed that it remains effective under such transformations, as shown in Table~\ref{tab: resist-quant}.
\subsection{Comparisons with More Fingerprinting}
\label{sec: more-defenses}
We have conducted comparisons with IF~\cite{xu2024instructional} and WLM~\cite{gu2022watermarking}, which are the most representative and widely adopted baselines. Other existing methods are largely variants of these two, with some performing even worse under our setting. To provide a more comprehensive evaluation against recent state-of-the-art approaches, we further include two of the most representative and up-to-date methods: EditMark~\cite{lieditmark} and PlugAE~\cite{yang2025challenge}.
As shown in Figure~\ref{fig:resistance-SOTA}, \sys achieves the highest \FSR while PlugAE and EidMark can be easily removed.
This is because they lack ``verification robustness'' and they depend on either special tokens or they depend on jointly multiple questions, thereby a single question removal can remove the whole fingerprint.

\begin{table}[t]
\centering
\small
\resizebox{\linewidth}{!}{
\begin{tabular}{l|l|l|l}
\toprule
Dataset & Domain & Language Style & Avg. Length \\
\midrule
AG News & News & Formal, headlines & $\sim$40 words \\
CNN & Articles & Formal, descriptive & $\sim$800 words \\
Daily & Conversations & Informal, multi-turn & 8--10 turns \\
ArXiv & Papers & Formal, technical & 150--250 words \\
\bottomrule
\end{tabular}
}
\caption{Descriptions of different datasets.}
\label{tab:dataset-summary}
\end{table}
\begin{table}[t]
\centering
\resizebox{\linewidth}{!}{
\begin{tabular}{@{}l|c|c|c|c@{}}
\toprule
    {Dataset} & 
    LLaMA2-7B & 
    LLaMA2-13B& 
    Mistral-7B & 
    Amber-7B \\
    \midrule
    AG News & 56\% & 59\% & 55\% & 53\% \\
    CNN & 57\% & 56\% & 56\% & 52\% \\
    Daily & 56\% & 60\% & 57\% & 53\% \\
    Arxiv & 56\% & 58\% & 55\% & 53\% \\
\bottomrule
\end{tabular}
}
\caption{Harmlessness of \sys on different datasets.}
\vspace{-0.2in}
\label{tab:harmless_sft_dataset}
\end{table}
\begin{table*}[t]
\centering
\begin{tabular}{|l|l|l|l|}
\hline
\textbf{Component} & \textbf{Linear} & \textbf{CNN} & \textbf{Attention} \\
\hline
Core Operation & $\mathbf{W} \mathbf{x}$ & $\text{Conv1d}(\mathbf{x})$ & $\text{Attention}(\mathbf{Q}, \mathbf{K}, \mathbf{V})$ \\
\hline
Layer Structure & $\mathbf{x} + \mathbf{W} \mathbf{x}$ & $\mathbf{x} + \text{GELU}(\mathbf{W} * \mathbf{x})$ & $\mathbf{x} + \mathbf{W}_o \text{Attn}(\mathbf{x})$ \\
\hline
Nonlinearity & None & GELU & Softmax \\
\hline
Weight Matrix & $\mathbf{W} \in \mathbb{R}^{d \times d}$ & $\mathbf{W} \in \mathbb{R}^{d \times d \times 3}$ & $\mathbf{W}_q, \mathbf{W}_k, \mathbf{W}_v, \mathbf{W}_o \in \mathbb{R}^{d \times d}$ \\
\hline
Parameters & $d^2$ & $3d^2$ & $4d^2$ \\
\hline
Computation & $O(d^2)$ & $O(d^2 \cdot L)$ & $O(L^2 \cdot d)$ \\
\hline
\end{tabular}
\caption{Encoder $E$ Architecture Comparison.}
\label{tab:model_comparison}
\end{table*}
\begin{table}[t]
\centering
\resizebox{\linewidth}{!}{
\begin{tabular}{lcc}
\toprule
\textbf{Methods} & \textbf{Effectiveness (FSR)} & \textbf{Harmlessness (GLUE)} \\
\midrule
IF (Xu et al. 2024a) & 100\% & 54\% \\
\textit{iSeal} & 100\% & 66\% \\
\bottomrule
\end{tabular}}
\caption{Effectiveness and harmlessness of \textit{iSeal} on LLaMa3-7B (AI 2024).}
\label{tab:llama3}
\end{table}
\begin{figure}[t]
    \centering
    \includegraphics[width=\linewidth]{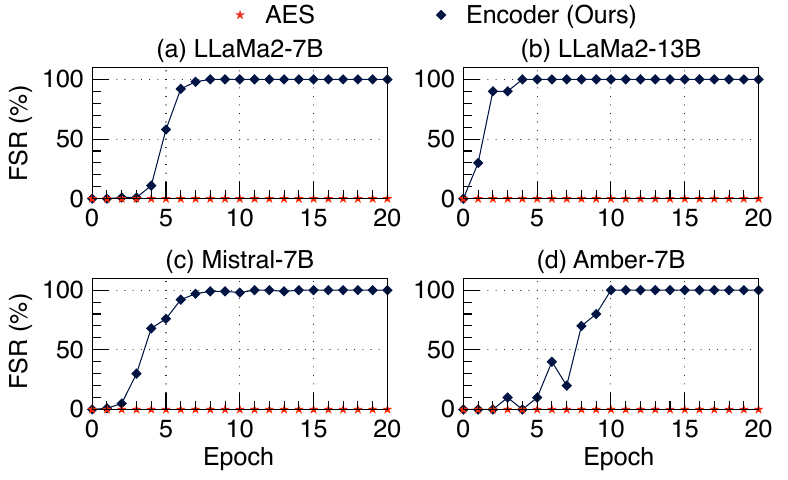}
    \caption{Impact of encryption method choices on the effectiveness of \sys. }
    \label{fig: abl-encoder}
\end{figure}
\begin{figure}[t]
    \centering
    \includegraphics[width=\linewidth]{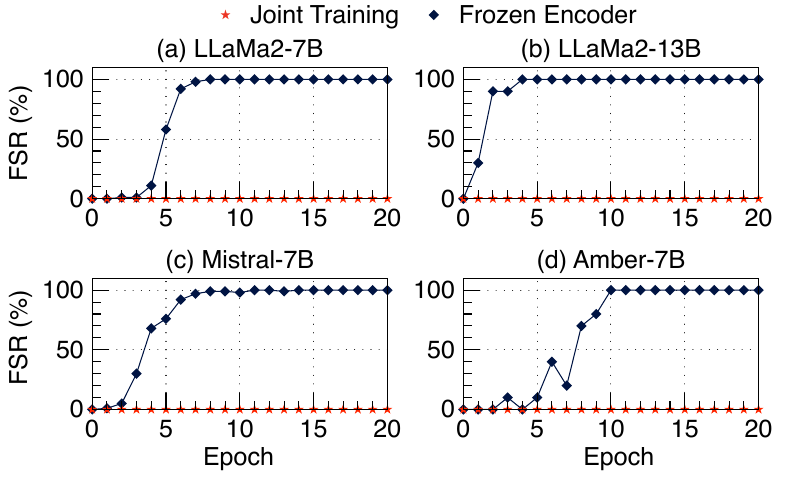}
    \caption{Impact of freezing the encoder on the effectiveness of \sys. }
    \label{fig: abl-freeze}
    \vspace{-0.2in}
\end{figure}
\begin{figure}[t]
    \centering
    \includegraphics[width=\linewidth]{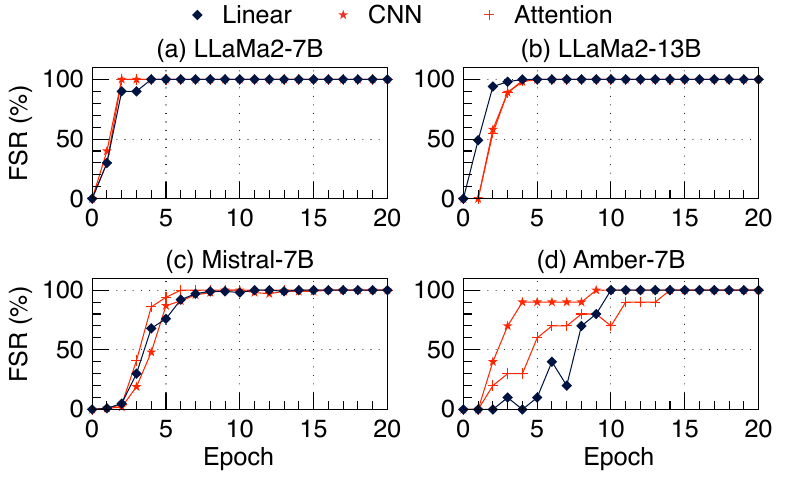}
    \caption{Impact of different structures on the effectiveness of \sys. }
    \label{fig: different-struct}
\end{figure}
\begin{figure}[t]
    \centering
    \includegraphics[width=\linewidth]{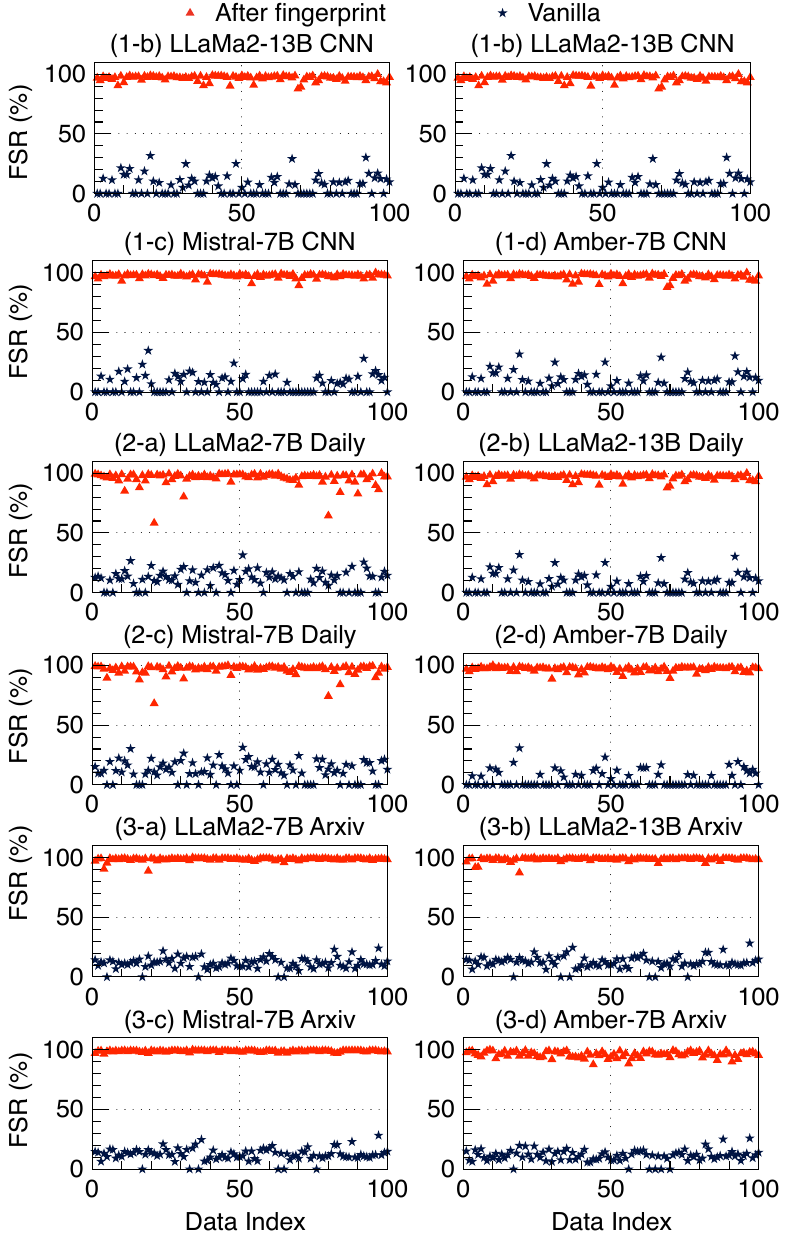}
    \caption{Effectiveness(\%) of \sys on different datasets. }
    \label{fig: different-data-sample}
    \vspace{-0.2in}
\end{figure}
\subsection{Performance on Different Settings}
\label{sec: different-settings}
\paragraph{Different Datasets.}
To demonstrate that \sys is both effective and harmless across diverse plaintext datasets, 
we further evaluate it on a variety of datasets with distinct linguistic and structural characteristics. 
As shown in Table~\ref{tab:dataset-summary}, we select datasets with varying input lengths, discourse styles, and topical domains, 
including AG News~\cite{zhang2015character}, CNN/DailyMail~\cite{li2017dailydialog}, DailyDialog~\cite{li2017dailydialog}, and arXiv Abstracts~\cite{clement2019arxiv}. 
These datasets collectively cover classification, summarization, and multi-turn dialogue tasks, enabling a comprehensive assessment of \sys under different conditions. 
The results in Figure~\ref{fig: different-data-sample} and Table~\ref{tab:harmless_sft_dataset} empirically support the effectiveness and harmlessness of our approach.

\paragraph{Different Models.}
%
To further validate the robustness and generalizability of \sys across model updates, we conduct a comprehensive evaluation on the recently released LLaMA3~\cite{meta2024llama3}-7B model—an architecture that reflects the latest advancements in large language model (LLM) design. This evaluation aims to examine whether the performance benefits of \sys persist when applied to stronger, more instruction-aligned models that may exhibit different sensitivities to fingerprinting interventions.
As summarized in Table~\ref{tab:llama3}, \sys maintains 100\% fingerprint success rate (FSR), matching the effectiveness of prior state-of-the-art methods such as IF~\cite{xu2024instructional}. However, \sys demonstrates a significantly higher GLUE score compared to the baseline, indicating superior harmlessness—i.e., less degradation in the model’s general-purpose capabilities after fingerprint injection. This result is particularly notable given the increased complexity and training stability of LLaMA3-7B, suggesting that \sys remains minimally invasive even when deployed on more recent and powerful foundation models.
The key to this advantage lies in the design principle of \sys: it avoids using natural language triggers or prompts as part of the fingerprinting mechanism. Unlike other methods that inject or fine-tune on human-readable phrases, thereby interfering with the model's linguistic distribution, \sys employs a latent-space approach that embeds the fingerprint at a representational level, abstracted away from semantic content. This reduces the likelihood that the fingerprint collides with downstream language tasks, leading to a more faithful preservation of the model’s original behavior.
In summary, the results on LLaMA3-7B reinforce the core claim of our work: \sys offers a fingerprinting solution that is not only effective across versions but also achieves a better trade-off between ownership verification and utility preservation than existing baselines.

\subsection{Ablation Studies \& Sensitivity Analysis}
\label{sec: ablation-studies}
\paragraph{Ablation Studies on the Training Strategy.}
To demonstrate the necessity of using a frozen encoder, we provide a comparison of learning curves from our ablation studies.
As shown in Figure~\ref{fig: abl-encoder} and Figure~\ref{fig: abl-freeze}, our method achieves optimal convergence within a few training rounds, which is consistent with our observations and claims in the main content.
This is because jointly training the encoder and decoder makes the base model and the fingerprinted model indistinguishable, leading to less stable training dynamics and making convergence more difficult.
The same issues occur for applying conventional cryptograph methods.

\paragraph{Sensitivity Analysis on the Encoder Architecture.}
To demonstrate that \sys can be applied to different encoder structures, we tested on two other representative structures for the encoder $E$, the details of which can be found in Table~\ref{tab:model_comparison} and the following definitions:
\begin{align*}
\text{Linear:} \quad &\mathbf{h}^{(l+1)} = \mathbf{h}^{(l)} + \mathbf{W}^{(l)} \mathbf{h}^{(l)} \\
\text{CNN:} \quad &\mathbf{h}^{(l+1)} = \mathbf{h}^{(l)} + \text{GELU}(\text{Conv1d}(\mathbf{h}^{(l)})) \\
\text{Attention:} \quad &\mathbf{h}^{(l+1)} = \mathbf{h}^{(l)} + \\&\mathbf{W}_o^{(l)} \text{softmax}\left(\frac{\mathbf{Q}^{(l)} {\mathbf{K}^{(l)}}^T}{\sqrt{d}}\right) \mathbf{V}^{(l)}.
\end{align*}
%
Figure~\ref{fig: different-struct} shows that \sys remains effective when applied to a variety of encoder structures, including both simple and more complex architectures. %
This demonstrates that \sys is flexible and does not rely on a specific network design to function correctly. %
Moreover, we find that using deeper or more expressive encoder structures can further enhance the cryptographic strength of the fingerprinting mechanism. %
This provides the model owner with an additional degree of freedom: by increasing the complexity of the encoder, they can make the fingerprint more entangled and harder to reverse engineer, thereby strengthening the overall security guarantees of \sys. %
These results suggest that \sys not only works robustly across different encoder configurations, but also benefits from structure-aware design choices made by the model owner.

\section{More Discussions}
\subsection{Reproducibility of \sys under Random Seeds}
To verify that the resistance of our method to unlearning is statistically significant, we evaluate it against three state-of-the-art unlearning attacks across four different models. Each experiment is repeated five times. As shown in Table~\ref{tab:fsr_unlearning}, we report both the mean and standard deviation. The results consistently demonstrate the effectiveness of our method under three unlearning scenarios.
This stability arises from the strong diffusion and confusion properties of our fingerprinting design. These properties ensure that it is difficult for an attacker to selectively unlearn the fingerprint without full access to the training configuration. In conclusion, our method exhibits both low randomness across trials and strong resistance to unlearning attempts.
\begin{table}[t]
\centering
\small
\begin{tabular}{l|cc|cc}
\toprule
\textbf{Model} & \multicolumn{2}{c|}{\textbf{Before Unlearning}} & \multicolumn{2}{c}{\textbf{After Unlearning}} \\
               & \textbf{Mean (\%)} & \textbf{Std (\%)} & \textbf{Mean (\%)} & \textbf{Std (\%)} \\
\midrule
LLaMa2-7B      & 100.0 & 0.00 & 99.8 & 0.4 \\
LLaMa2-13B     & 100.0 & 0.00 & 99.8  & 0.4 \\
Mistral-7B     & 100.0 & 0.00 & 99.8 & 0.4 \\
Amber-7B       & 100.0 & 0.00 & 99.8  & 0.4 \\
\bottomrule
\end{tabular}
\caption{\FSR before and after applying unlearning attacks. Results are averaged over 5 trials.}
\label{tab:fsr_unlearning}
\end{table}

\subsection{Robustness to Fine-Tuning on Hexstream}
Considering that our method embeds fingerprints using hexadecimal representations, we further evaluate its robustness under fine-tuning on a large hex-code dataset (a dataset composed of 10k random hexadecimal data for autoregressive learning). This experiment simulates a realistic threat scenario where an adversary attempts to overwrite or unlearn the fingerprint by continuing training on hex-style data.
We find that \sys maintains a $100\%$ \FSR even after extensive fine-tuning on this dataset. Importantly, the model's performance on general tasks remains stable, with no significant drop in zero-shot SuperGLUE scores. This demonstrates that our fingerprint remains intact and verifiable under domain-consistent fine-tuning that does not sacrifice the model’s utility.
This is because our method, like IF~\cite{xu2024instructional}, adopts adapter-based conditional training. Without knowing the specific condition used during fingerprint injection, an attacker cannot easily unlearn or remove the embedded fingerprint.

\subsection{Why Ownership Cannot Be Overclaimed}
\textit{First}, our fingerprinting framework inherently prevents ownership overclaim, a critical limitation in prior \textit{passive} fingerprinting approaches. Passive methods rely solely on extracting implicit features from pre-trained models without any active embedding process. As a result, they are vulnerable to ownership overclaim: any party with access to the model weights can falsely assert ownership, since no one is uniquely linked to the fingerprint. This is similar to using a permissively licensed GitHub repository (e.g., MIT license), where any user can replicate the project and claim authorship regardless of actual contribution.
In contrast, our method uses a \textit{proactive} fingerprinting strategy. During training, we inject a secret-derived fingerprint into the model using an external encoder initialized with a private key. Only the original trainer, who possesses the secret and performs the injection during training, can generate a valid ownership claim. This asymmetric design ensures that model thieves, who lack access to the training setup, cannot reproduce or recover the fingerprint afterward.

\textit{Moreover}, since the judge assigns the verification queries and responses, no adversary can falsely claim ownership by generating passive fingerprints.

\textit{Furthermore}, even if a third-party entity fine-tunes our fingerprinted model, they can only claim ownership of the fine-tuned variant, not the base model. This is because the fingerprint embedded in the original model remains intact and cannot be removed or replaced without full retraining. To support this, we added five new fingerprints using different keys and confirmed that the original fingerprint still remains detectable across four different LLMs (\ie, LLaMa2 7B \& 13B, Mistral 7B, and Amber-7B).
Moreover, since model thieves only have read access to the base model, they cannot manipulate it to inject their own fingerprints.
This property also provides an additional benefit: it allows fine-tuning contributors to claim credit for their own work, similar to how contributors to an open-source fork can be acknowledged without affecting the authorship of the original project.
This asymmetric claimability reinforces our framework’s ownership guarantee and aligns with collaborative and open-source development practices. The original model trainer retains exclusive rights to the base fingerprint, while fine-tuners can claim authorship of their own modifications without compromising the integrity of the fingerprinting mechanism.

\end{document}